\documentclass[12pt]{article}
\usepackage{geometry}

\usepackage{amsmath, amssymb, graphicx, fullpage, color, mathtools, amsthm, xcolor}
\usepackage{caption, subcaption}
\usepackage{algorithm, algorithmic}
\usepackage{cite}
\usepackage{subcaption}
\usepackage{array}
\usepackage{url}
\usepackage{tikz}
\usepackage{tabularx}
\usepackage{makecell}
\usepackage{bm}
\usepackage{microtype}
\usepackage{enumerate}
\usepackage{hyperref, color}
\usepackage{indentfirst}
\newcolumntype{C}[1]{>{\centering\let\newline\\\arraybackslash\hspace{0pt}}m{#1}}

\usepackage{xr}
\usepackage{cleveref}

\definecolor{webgreen}{rgb}{0,.35,0}
\definecolor{webbrown}{rgb}{.6,0,0}
\definecolor{RoyalBlue}{rgb}{0,0,0.9}
\definecolor{purp}{rgb}{0.6,0.05,0.8}
\definecolor{ora}{rgb}{0.7,0.35,0.02}

\hypersetup{
   colorlinks=true, linktocpage=true, 
   urlcolor=webbrown, linkcolor=RoyalBlue, citecolor=webgreen,
   pdfauthor={Jiani Fang, Xiaoxiao Ding, Gary P. T. Choi},
   pdfsubject={Geometric quantification for nonlinear deformation in knitted fabrics}
}

\title{Geometric quantification for nonlinear deformation in knitted fabrics}

\author{Jiani Fang$^{1}$, Xiaoxiao Ding$^{2,\ast}$, Gary P. T. Choi$^{1,\ast}$\\
\\
\footnotesize{$^{1}$Department of Mathematics, The Chinese University of Hong Kong, Hong Kong, China}\\
\footnotesize{$^{2}$School of Engineering and Applied Sciences, Harvard University, Cambridge, MA 02138, USA}\\
\footnotesize{$^\ast$To whom correspondence should be addressed; E-mail: xding@alumni.harvard.edu, ptchoi@cuhk.edu.hk}
}
\date{ }

\begin{document}

\maketitle

\begin{abstract}
\noindent Knitted fabrics exemplify a broad class of architected materials capable of large deformations, enabling shape morphing, mechanical biocompatibility, and embedded multifunctionality without material damage. Although geometric nonlinearity has been intuitively utilized in their design, a quantitative description of stitch-resolved deformation and its temporal evolution remains lacking. Here, we introduce a geometric quantification framework that reconstructs smooth yarn centerlines and fabric surfaces from sparse yarn-level representations and extracts interpretable descriptors across dimensions. Applied to representative knitted structures, this framework resolves how global deformation is distributed among stitch reorientation, loop bending, surface bending, and dilation. Moreover, it reveals how regions of large geometric variation emerge, persist, and redistribute over time. Rather than directly measuring stress, these geometric descriptors define a unified geometric state space for comparing knitted structures and identifying candidate regions of mechanical localization. The framework provides a quantitative language for nonlinear deformation in knits and establishes a geometry-based representation that can be coupled to constitutive models, experimental measurements, and graph-based inverse-design workflows.
\end{abstract}


\pagebreak

%%%%%%%%%%%%%%%%%%%%%%%%%%%%%%%%%%%%%%%%%%%%%%%%%%%%%%%%%%%%%%%%%%%%%%%%%%%%%%%%%%%%%%%%%%%%%%%%%%%%%%%%%%%%%%%%%%%%%%%%%%%%%%%%%%%%%%%%%%%%%%%%%
\section{Introduction}
Architected materials whose mechanical behavior arises from geometry rather than composition~\cite {bertoldi2017flexible} have opened new possibilities for designing extremely deformable~\cite{Moestopo2023, surjadi2025double, pescialli2025topology}, shape morphing~\cite{choi2019programming, choi2021compact, hong2022boundary, dudek2025shape}, and multifunctional systems~\cite{el2021digital, mohammadi2021flexible, jiao2023mechanical}. Among these, knitted fabrics, an everyday object yet complex multiscale system, stand out as structural candidates for soft robots~\cite{luo2022digital, sanchez20233d, wang2024soft}, wearable sensors~\cite{luo2021, wan2024all, kou2024wearable, zhou2024highly}, and biomimetic devices~\cite{han2017blooming, ali2017knitting, FU2022137241}. Their ability to undergo large, reversible, and anisotropic deformations while maintaining structural integrity makes them promising. A quantitative understanding of how knitted fabrics respond to external loading through localized shape, deformation, and redistribution over space and time is essential to guide inverse design. Prior work has developed topology-level representations for knitted textiles~\cite{paras2020geometric, kapllani2021topoknit,kapllani2022loop}, providing a foundation for assembly language and property analysis. However, a general framework for quantifying how such knitted geometries deform locally and evolve stitch by stitch under loading remains limited, particularly as knitted materials exhibit geometrically complex deformation processes~\cite{rout2022,niu2025geometric}. 

Yarn-based simulations~\cite{bergou2008discrete, Kaldor2010, ding2024unravelling} have provided powerful computational tools to probe into these processes at the micro-mechanical level. Ding et al.~\cite{ding2024unravelling} tracked both the alignment of the yarn segments with the loading direction and the stretching of the individual yarn segments, which offered insights into the evolution of localized stress concentrations. Experimental techniques for tracking mechanical hot spots, mainly optical tracking methods, have been widely adopted for surface strain analysis~\cite{Vic2D, DICengine}. However, their reliance on image texture limits accurate quantification of textiles, which are often challenged by self-occlusion and significant out-of-plane deformation caused by fabric boundary warping. Conversely, micro-computed tomography~\cite{ROUX20081253} provides detailed microstructural data, yet they tend to be resource-intensive and restricted to small-scale samples.

\begin{figure}[h!]
    \centering
    \includegraphics[width=0.95\linewidth]{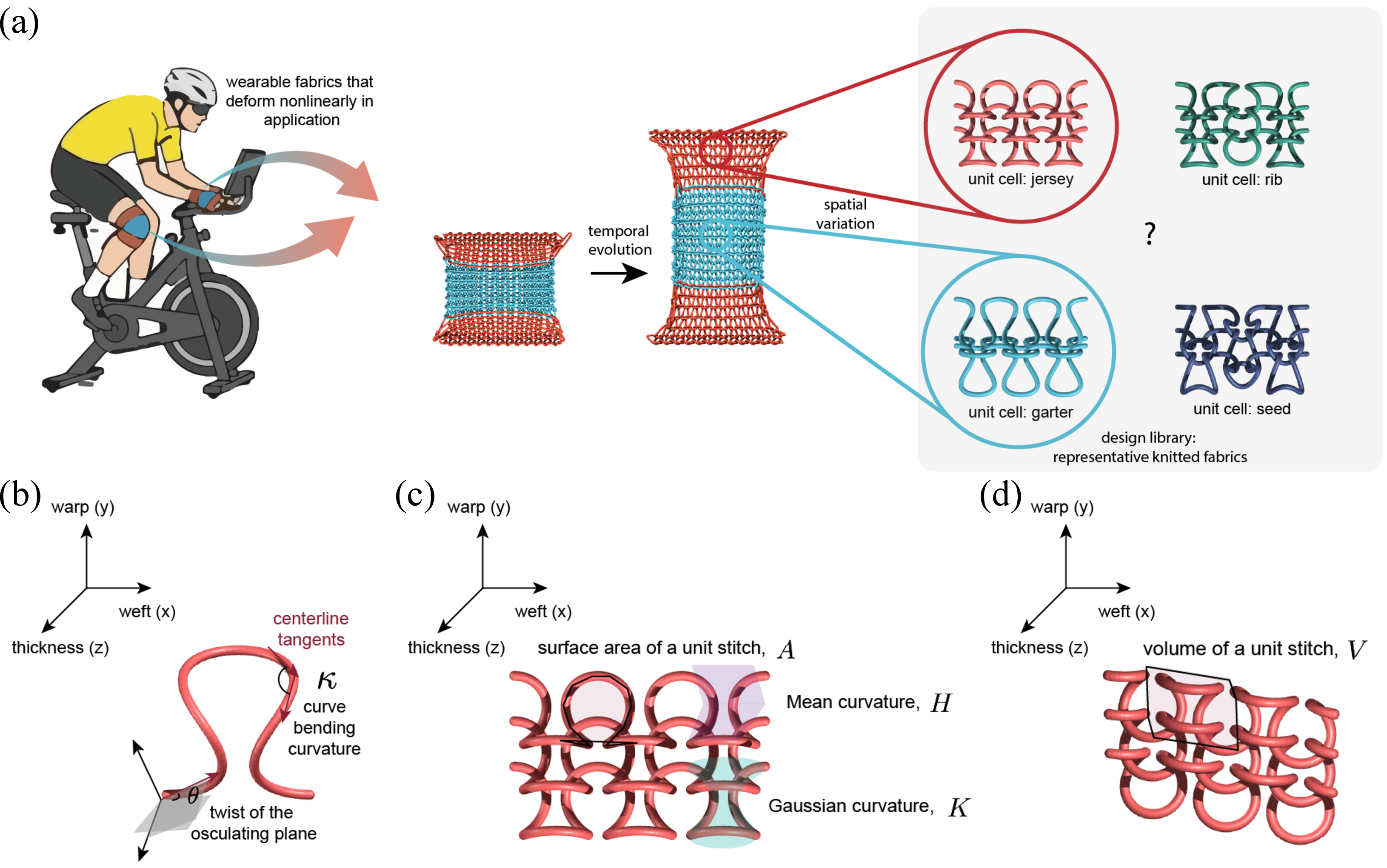}
    \caption{\textbf{An illustration of our geometric quantification framework.} (a) Conceptual workflow motivating the quantification of localized mechanical responses in hybrid fabrics, such as guiding the inverse design of an optimally conformable patch for cyclists during training sessions. A knitted sleeve undergoes temporal evolution under large deformation while exhibiting spatial variation represented by local constitutive responses. Representative unit cells (RUC) selected from the design library of knitted fabrics consisting of jersey, garter, rib, and seed. The geometric models and deformation data are adapted from~\cite{ding2024unravelling}. (b)~Descriptors of a unit stitch centerline including the local bending curvature ($\kappa$), which measures how much the centerline curve is bending in the osculating plane, and the torsion ($\tau$), which measures how much the centerline curve is twisting out of the osculating plane. Here, $\tau = d \theta / ds$, where $s$ is the arc length and $\theta(s)$ is the angle of rotation of the osculating plane around the tangent line. (c)~Illustration of the surface curvatures (mean curvature, $H$, and Gaussian curvature, $K$) of a unit stitch. (d)~Illustration of the volume of a unit stitch.}
    \label{fig:illustration}
\end{figure}

In this work, we develop a geometric quantification framework for measuring and analyzing the deformation of knitted fabric structures (Fig.~\ref{fig:illustration}) stitch by stitch. From this representation, we extract non-affine metrics that quantify stitch reorientation, anisotropic stretching, and out-of-plane bending, geometric features that are often otherwise homogenized. By abstracting the geometry across dimensions, our approach bridges the gap between yarn-scale and fabric-scale parameters. It enables the identification and tracking of geometric hot spots, namely regions of persistent localized variation, and provides a basis for inferring stress and energy localization when coupled with constitutive modeling. We demonstrate the framework on four canonical knitting patterns, two mixed-pattern fabrics, and a cylindrical knitted actuator. Compared with texture-dependent approaches, this framework yields continuous geometric fields even in regions affected by weak texture, self-occlusion, or out-of-plane motion. Because the resulting descriptors are defined on stitches and their connectivity, the framework is naturally compatible with graph-based learning for inverse design and damage analysis~\cite{paul2022graphbased, karapiperis2023prediction, zheng2023unifying}. Meanwhile, our framework can achieve stitch-level curve reconstruction and geometric quantification of fabric and fiber structures, and help reveal the geometric mechanism of how structural parameters affect macroscopic properties, thereby facilitating relevant research in geometrically programmed architectures~\cite{huang2024bending,yang2026shape}. Beyond textiles, our framework offers a generalizable methodology for tracking dynamic deformation and reversible recovery processes of reconfigurable metamaterials~\cite{bai20263d} and quantifying heterogeneous deformation in other flexible systems~\cite{patil2020topological,TONG2026102430}, where complex shape changes challenge traditional measurement approaches. Moreover, the geometric features extracted from our framework provide reliable validation benchmarks for model training, which can be effectively used for data-driven structural design~\cite{yao2025experimental}.

%%%%%%%%%%%%%%%%%%%%%%%%%%%%%%%%%%%%%%%%%%%%%%%%%%%%%%%%%%%%%%%%%%%%%%%%%%%%%%%%%%%%%%%%%%%%%%%%%%%%%%%%%%%%%%%%%%%%%%%%%%%%%%%%%%%%%%%%%%%%%%%%%
\section{Results}
The quantification methods used to generate spatial and temporal analyses presented in this work are fully detailed in the Methods section. Simulated deformation data from uniaxially stretching a total of six representative 2.5D fabrics (adapted from~\cite{ding2024unravelling}) and bending a 3D knitted actuator are first reconstructed into a generalized two-dimensional matrix; then stitch-by-stitch descriptors are computed and visualized as heat maps showing spatial information and line graphs showing temporal information. The boundary layers that were tethered during simulation, specifically the top two and bottom two layers, were excluded to minimize boundary effects. To avoid cancellation, all averages in the following sections are taken over absolute values.

%%%%%%%%%%%%%%%%%%%%%%%%%
\subsection{Spatial variation of 2.5D fabrics}

\begin{figure}[t]
    \centering
     \includegraphics[width=1\linewidth]{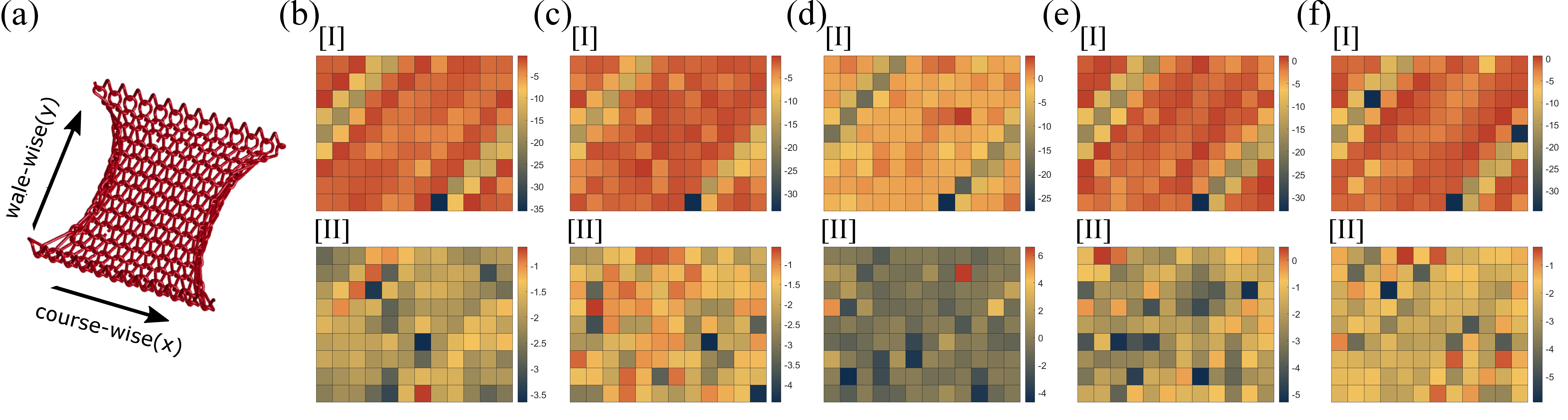}
    \caption{\textbf{Quantification of the anisotropic mechanical responses in the jersey pattern.} (a) The jersey pattern. (b)--(f) Spatial distribution of temporal changes in five representative quantities under tensile strain from 0\% to 120\%, where the color of each cell denotes the total change in the geometric quantity for the corresponding stitch: (b) Curve curvature. (c) Curve torsion. (d) Gaussian curvature. (e) Area. (f)~Volume. Loading directions: [I]~Course‐wise tension. [II]~Wale-wise tension.}
     \label{fig2}
\end{figure}

\subsubsection{Anisotropic responses}
Across all four representative knitted patterns (jersey, garter, rib, seed), we observe pronounced anisotropy that is naturally captured as directional bias. Under uniaxial loading, the manifold-based fields reveal that deformation is not uniformly accommodated by stretch (shown as area and volume changes) alone: instead, it partitions into a combination of (i)~stitch reorientation characterized by torsion $\tau$, (ii)~localized curve bending and straightening quantified by curve curvature $\kappa$, and (iii)~surface bending jointly quantified by surface curvatures (Gaussian curvature $K$ and mean curvature $H$). The relative dominance of these mechanisms depends on knitted structure topology and loading direction.

In general, when fabrics are subjected to course‐wise tension, we see stripe-like spatial gradients appearing regularly in all geometric descriptors, showing simultaneous responses from stitch reorientation, localized curve bending, and surface bending (Fig.~\ref{fig2}(b)--(f): [I]). By contrast, when fabrics are subjected to wale-wise tension, the overall changes are more complicated. As illustrated in Fig.~\ref{fig2}(b)--(f): [II], there lacks a shared general trend across all descriptors, showing the interplay of multiple mechanisms. Curve torsion seems to have the most varied spatial distribution, indicating that stitch reorientation dominates the overall mechanical response. 

Importantly, these anisotropic signatures emerge directly from geometric observables defined on the manifold, providing a pattern-agnostic way (results on garter, rib, and seed in SI section S4 and Fig.~S10--S12) to compare knit architectures using a common descriptor set.

\begin{figure}[t!]
     \centering
     \includegraphics[width=1\linewidth]{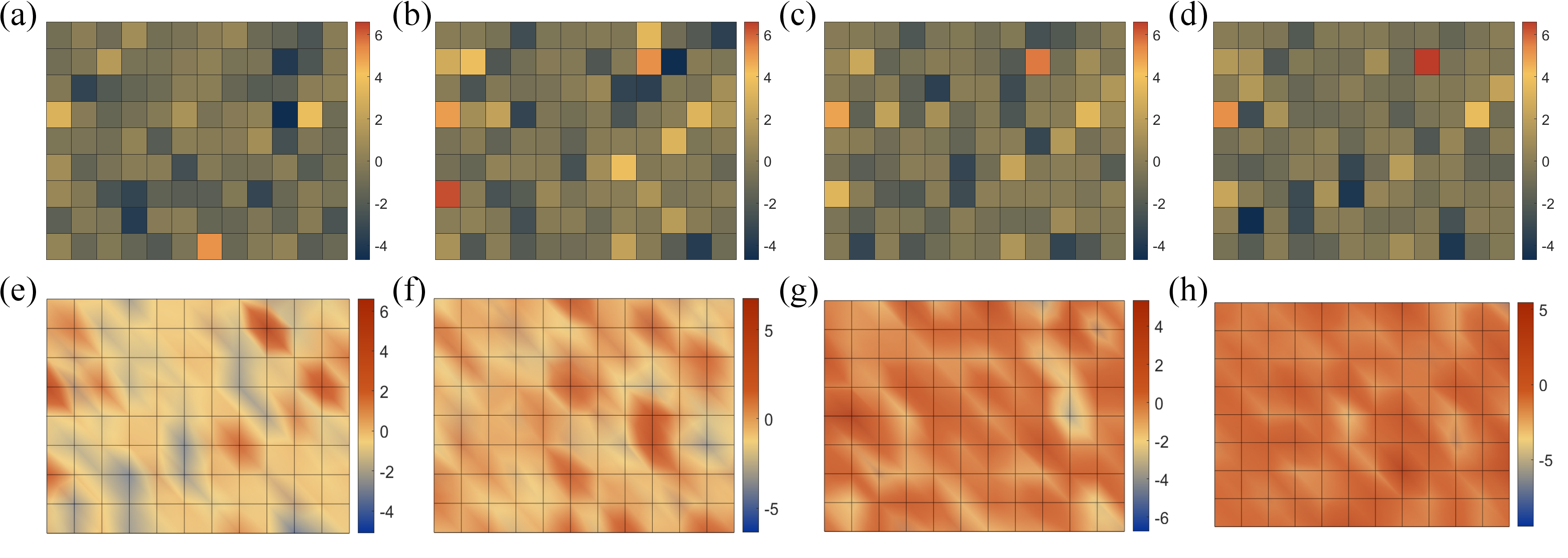}
     \caption{\textbf{Heterogeneous mechanical responses and hot spots of Gaussian curvature under wale-wise tension.} (a)--(d) Gaussian curvature change in jersey fabric with 20\%, 60\%, 80\%, and 120\% wale-wise tensile strain. (e)~Time-aggregated maps of changes in Gaussian curvature for jersey fabric, generated by overlaying normalized heat maps acquired during successive tensile loading at 0\%, 20\%, 40\%, 60\%, 80\%, 100\%, and 120\% strain with applied spatial smoothing. (f)~Time-aggregated maps of changes in Gaussian curvature for garter fabric. (g)~Time-aggregated maps of changes in Gaussian curvature for rib fabric. (h)~Time-aggregated maps of changes in Gaussian curvature for seed fabric. }
     \label{fig3}
 \end{figure}

%%%%%%%%%%%%%%%%%%%%%%%%%%%%
\subsubsection{Heterogeneous responses}
Beyond global anisotropy, the fabrics exhibit strong spatial heterogeneity, particularly when they are loaded wale-wise. Fig.~\ref{fig3}(a)--(d) show through the example of a jersey pattern that large variations in Gaussian curvature, $K$, localize into spatially concentrated regions. We refer to these regions as geometric hot spots, namely stitch neighborhoods that sustain large absolute geometric variation over multiple loading increments. 

To visualize persistence, in Fig.~\ref{fig3}(e)--(f) we overlay the normalized heat maps from successive loading steps and apply spatial smoothing to obtain time-aggregated maps of large $|\Delta K|$ for all four representative patterns (jersey, garter, rib, seed). The normalized matrix for generating the heatmap is derived by normalizing the change in each quantity relative to the initial frame, defined as $ \Delta Q _t = Q_t - Q_0$. It is then divided by the initial value to yield the percentage change relative to the initial state, expressed by the formula $P_t=\frac{\Delta Q_t}{Q_0}$. The resulting fields do not directly measure stress or stored energy; instead, they identify candidate regions of geometric localization that persist throughout deformation. Across patterns, these persistent regions remain spatially structured rather than random, indicating that localization is constrained by stitch topology and connectivity (see also SI Section~S4 and Fig.~S13--S16).

This geometric viewpoint is useful for two reasons. First, it reveals where deformation is repeatedly concentrated, thus predicting potential structural vulnerabilities within the material. Second, it provides a common language for comparing how different knit architectures redistribute deformation.

\begin{figure}[t!]
    \centering
    \includegraphics[width=1\linewidth]{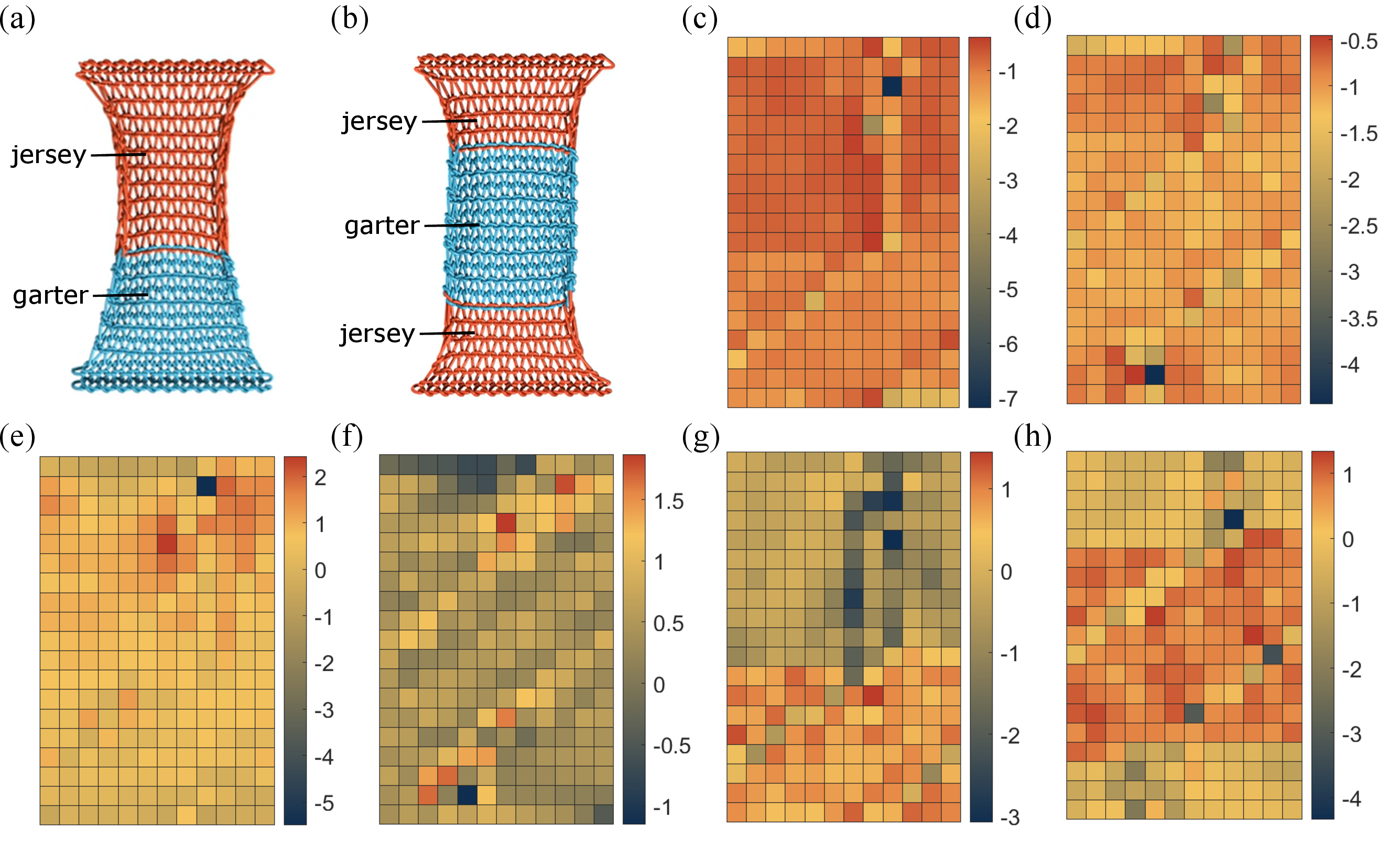}
    \caption{\textbf{Quantifying the mechanical responses from mixed-pattern fabrics.} Spatial distribution of geometric quantity changes in two mixed-pattern fabrics under wale-wise uniaxial tensile strain from 0\% to 180\%. (a) Visualization for pattern A. (b)~Visualization for pattern B. (c) Curve curvature change for pattern A. (d) Curve curvature change for pattern B. (e) Aspect ratio change for pattern A. (f) Aspect ratio change for pattern B. (g) Volume change for pattern A. (h) Volume change for pattern B.}
    \label{fig4}
\end{figure}

%%%%%%%%%%%%%%%%%%%%%%%%%%%%%
\subsubsection{Responses from mixed-pattern fabrics}

Besides considering the four representative knitted patterns (jersey, garter, rib, seed) separately, it is also natural to analyze fabrics composed of multiple patterns. Here we consider two mixed-pattern fabrics made of jersey and garter (Fig.~\ref{fig4}): 

\begin{itemize}
    \item Pattern A (Fig.~\ref{fig4}(a)): A mixed pattern with the top half being jersey and the bottom half being garter. 
    \item Pattern B (Fig.~\ref{fig4}(b)): A mixed pattern with the top and bottom quarter being jersey and the middle half being garter.
\end{itemize}

Figure~\ref{fig4}(c)--(h) shows that the mixed fabrics do not respond as a simple spatial average of their constituents (see also SI Section~S4 and Fig.~S17). Instead, each stitch type retains a characteristic response signature while the interfaces between stitch domains reorganize the overall field. In both patterns, the garter regions display oblique bands of geometric activity and comparatively stronger variation in volume, whereas the jersey regions exhibit a more spatially uniform response together with larger changes in curve curvature. This contrast is consistent with a stronger contribution from dilation and rearrangement in garter-dominated domains, and a stronger contribution from localized loop bending and straightening in jersey-dominated domains.

A second notable feature is the reproducibility of the response within repeated domains of the same stitch type. In Pattern~B (Fig.~\ref{fig4}(b)), for example, the two outer jersey regions exhibit qualitatively similar distributions of curvature, aspect ratio, and volume variation despite being separated by a garter region. This preservation of stitch-specific response signatures suggests that the proposed descriptor set can separate intrinsic pattern behavior from the additional localization induced by pattern interfaces.

These observations are relevant for design. Mixed-pattern fabrics are often used to program heterogeneous compliance, fit, and actuation, and the present framework resolves not only where deformation is concentrated, but also which geometric mode dominates in each subdomain. This provides a practical means to compare candidate multi-pattern layouts before coupling them into a full mechanical or inverse-design pipeline.

%%%%%%%%%%%%%%%%%%%%%%%%%%%%%%
\subsection{Temporal change of the spatial variation of 2.5D fabrics: Onset and evolution of geometric hot spots}
\begin{figure}[t]
     \centering
     \includegraphics[width=1\linewidth]{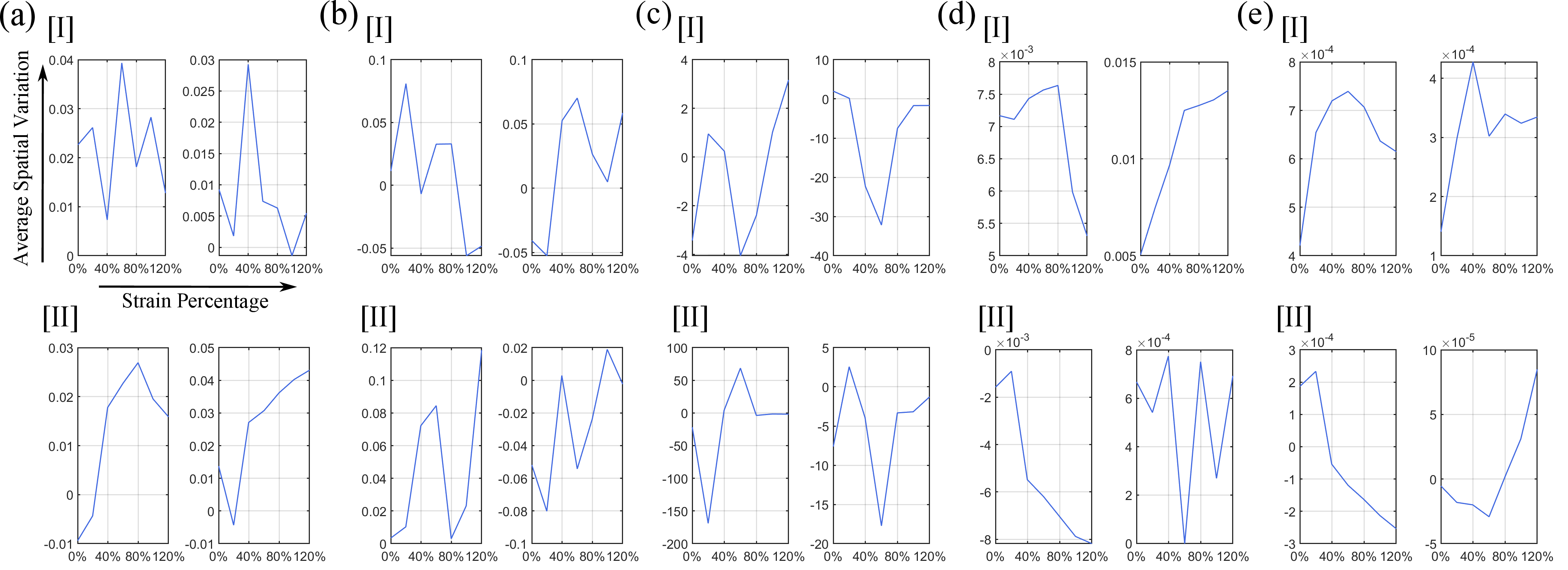}
     \caption{\textbf{Temporal changes in spatial variations of four representative quantities for jersey fabric, recorded across a tensile strain range of 0\% to 120\%.} (a)~Curve curvature, $\kappa$. (b)~Curve torsion, $\tau$. (c) Gaussian curvature, $K$. (d) Area. (e) Volume. Loading directions: [I]~Course-wise. [II]~Wale-wise. In each of (a)--(e), the left plot corresponds to course-wise variation and the right plot corresponds to wale-wise variation.}
     \label{fig5}
 \end{figure}

We next characterize how mechanical responses evolve over time under monotonically increasing uniaxial load. The line plots in Fig.~\ref{fig5} summarize the course‐wise and wale-wise evolution of $\kappa$, $\tau$, $K$, area, and volume for a jersey knit, respectively (see SI Section S4 and Fig.~S18--S20 for garter, rib, and seed knits).

Two trends are evident. First, the temporal evolution is descriptor dependent: some quantities vary smoothly with load, whereas others exhibit non-monotone behavior, indicating redistribution between reorientation, bending, and dilation as deformation proceeds. Second, the evolution is loading-direction dependent. In jersey, course-wise loading produces comparatively regular trends across descriptors, consistent with the stripe-like spatial patterns previously observed in Fig.~\ref{fig2}, whereas wale-wise loading generates more irregular temporal signatures, reflecting stronger competition among local deformation modes.

Taken together with the heat maps, these curves suggest that geometric localization develops progressively rather than appearing instantaneously. Early loading is accompanied by broadly distributed reorientation, after which localized bending and surface-shape changes become more pronounced at selected stitch neighborhoods. We therefore interpret the temporal response as an onset-and-redistribution process: load is initially accommodated by diffuse geometric rearrangement and subsequently reorganized into more spatially heterogeneous descriptor fields. In this sense, when localization emerges is as important as where it emerges, and both require stitch-resolved geometric tracking.

%%%%%%%%%%%%%%%%%%%%%%%%%%%%
\subsection{Spatial and temporal variations of a 3D knit}
\begin{figure}[t]
        \centering 
         \includegraphics[width=\linewidth]{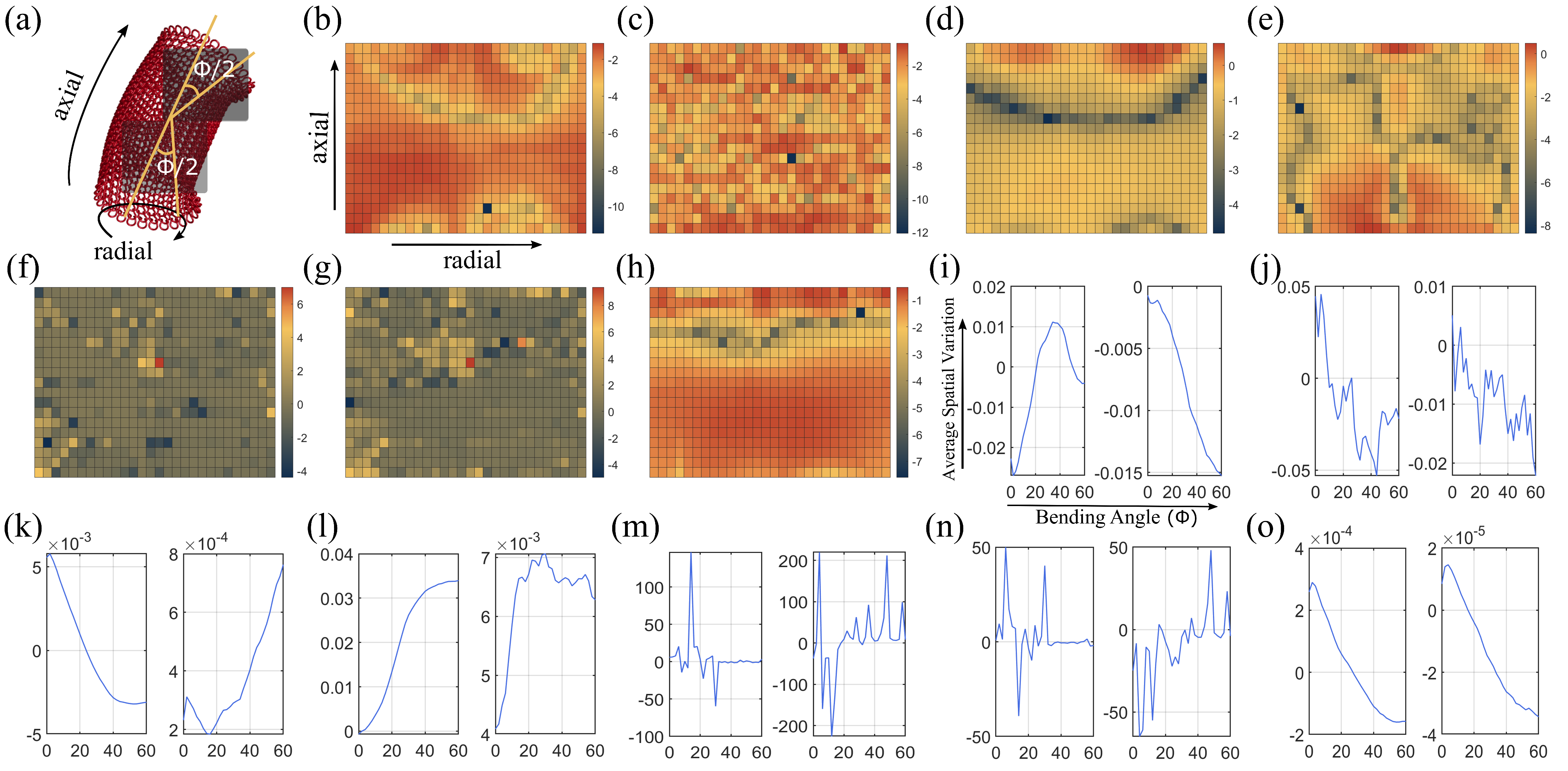}
         \caption{\textbf{Quantification methods applied on the deformed cylinder sample.} (a)~The deformed cylinder sample. (b)--(h)~Spatial distribution of change in seven representative quantities as the bending angle increases from $0^{\circ}$ to $60^{\circ}$. (b)~ Curve curvature. (c)~Curve torsion. (d)~Area. (e)~Aspect ratio. (f)~Gaussian curvature. (g)~Mean curvature. (h)~Volume. (i)--(o)~Temporal changes of spatial variation in seven representative quantities as the bending angle increases from $0^{\circ}$ to $60^{\circ}$. (i)~Curve curvature. (j)~Curve torsion. (k)~Area. (l)~Aspect ratio. (m)~Gaussian curvature. (n)~Mean curvature. (o)~Volume. In each of (i)--(o), the left plot corresponds to variation along the radial direction and the right plot corresponds to variation along the axial direction.}
        \label{fig6}
\end{figure}

Finally, we extend the framework from planar fabrics to a 3D cylindrical knitted structure that is typically deployed as a pneumatic actuator. In Fig.~\ref{fig6}, we summarize the analysis using the aforementioned geometric quantification methods. Here, we focus on the shape changes of the structure along the radial and axial directions (Fig.~\ref{fig6}(a)). This 3D knit exhibits pronounced regionalization: spatial maps (Fig.~\ref{fig6}(b)--(h)) reveal zones dominated by bending (reflected in high $\kappa$ values (b)), zones dominated by anisotropic stretching (reflected by aspect ratio change (e)), and zones dominated by dilation (reflected in distinctive stitch area and volume increases in (d) and (h)), with transitions that correlate with global geometric features. 

As for the temporal evolution, Fig.~\ref{fig6}(i)--(o) reveal a consistent descriptor-dependent pattern: quantities including curve curvature, area, aspect ratio, and volume exhibit smooth, monotonic variations throughout loading, while curve torsion, Gaussian curvature, and mean curvature display distinct fluctuating behavior. Notably, spatial variations of these quantities along the axial direction stabilize by the end of the bending process, which can be modeled using the previous results of 2.5D fabric. Here, each axial strip can be treated as an element undergoing uniaxial stretching (with an extra bending force), exhibiting a clear two-stage deformation process: [I]: curve reorientation, bending, and straightening, with surface bending. [II]: yarn dilation with surface bending. In contrast, radial spatial variations are far more complex and do not reach a stable state in most quantities, as each circular segment undergoes partial expansion at the outer edge and partial contraction at the inner edge. 

%%%%%%%%%%%%%%%%%%%%%%%%%%%%%%%%%%%%%%%%%%%%%%%%%%%%%%%%%%%%%%%%%%%%%%%%%%%%%%%%%%%%%%%%%%%%%%%%%%%%%%%%%%%%%%%%%%%%%%%%%%%%%%%%%%%%%%%%%%%%%%%%%
%V3.
\section{Discussion}

The central contribution of this work is a stitch-resolved geometric framework for representing nonlinear deformation in knitted fabrics through a small set of interpretable local descriptors. Rather than reducing the response to a single homogenized strain field, the framework resolves how global loading is partitioned among reorientation, bending, surface reconfiguration, and dilation. This decomposition is useful because different knit architectures may accommodate the same external loading through different local geometric mechanisms.

The results on four representative knitted patterns show that both anisotropy and heterogeneity emerge naturally when these descriptors are analyzed at stitch resolution. The mixed-pattern examples further show that stitch-type-specific response signatures persist inside multi-pattern layouts, while interfaces between stitch domains introduce additional localization. The extension to a 3D knitted actuator highlights an additional coupling between local stitch topology and global embedding geometry: bending, aspect-ratio change, and dilation become regionally biased when the same stitch architecture is placed on a curved body. 

An important point is interpretive scope. The quantities computed here are geometric observables. They identify regions of persistent geometric concentration and therefore provide candidate sites of mechanical localization. For instance, at the yarn level, consider an infinitesimal yarn segment with total elastic energy defined as
\begin{equation}
    E_{\text{total}} = E_s + E_b + E_t,
\end{equation}
where the tensile, bending, and torsional potential energies follow 
\begin{equation}
    E_s = \frac{1}{2}EA \int \epsilon_{\text{local}}^2(s) \mathrm{d}s, \quad 
    E_b = \frac{1}{2}EI \int \kappa^2(s) \mathrm{d}s, \quad 
    E_t = \frac{1}{2} GJ \int \tau^2(s) \mathrm{d}s.
\end{equation}
Here, $E$ is the Young's Modulus, $A$ is the cross-sectional area, $\epsilon_{\text{local}}$ is the local axial strain, $I$ is the second moment of inertia, $G$ is the shear modulus, and $J$ is the torsional constant. Therefore, changes in curve curvature and torsion directly map to the redistribution of yarn segment elastic energy. At the surface level, our quantification of the area variation can be effectively used for interpreting the accumulation of in-plane strain localization via 
\begin{equation}
    \Delta A \propto \varepsilon_1 + \varepsilon_2,
\end{equation}
where $\Delta A$ is the area change rate, and $\varepsilon_1$ and $\varepsilon_2$ are the in-plane principal strains. At the spatial scale, abnormal unit volume variations quantified by our method can directly capture pre-instability stitch structural distortions, including inter-yarn slippage-induced relaxation, stitch torsion, and folding, which disrupt the ordered fabric architecture and alter inter-stitch constraints.

Establishing such constitutive links will require coupling the present geometric state space to yarn mechanics, contact, friction, and dissipation, or validating the descriptors against experimental full-field measurements and failure data. A feasible strategy is to compare the candidate localization regions identified by geometric descriptors with experimentally measured full-field mechanical localization maps that precisely capture the occurrence of strain concentration and instability. The geometric-mechanical correspondence and the early-warning timeliness of geometric indicators can then be quantitatively verified by evaluating both the spatial overlap ratio and the onset time difference between the two sets of regions. We view the present framework as a quantitative geometric layer that makes such future couplings possible.

This perspective nonetheless has immediate practical value with various applicable scenarios. Leveraging its cross-topology applicability, precise differentiation of geometric properties across distinct knitting regions, and stable tracking of structural evolution at interfaces, this quantification framework enables effective detection of localized mechanical regions in non-uniform knitted topologies. Regarding textile structures, the curve reconstruction method also applies to multi-layer fabrics, with additional labels to distinguish nodes on distinct yarns. For time scale selection, although the proposed framework is developed from quasi-static equilibrium data, its core methodology is not constrained by the loading type. For instance, it is fully compatible with cyclic loading conditions in the quasi-static regime and is expected to be more advantageous than optical-based methods, which may be affected by occlusion. The framework may also be applied to dynamic loading scenarios given high-frequency sampling data~\cite{RADI2023104658}. Moreover, the curve reconstruction method cuts the number of nodes required to capture yarn structures, enabling a sparse node displacement tracking strategy that reduces the single-frame data volume during acquisition. Also, because the representation is defined on stitches and their connectivity, it is naturally compatible with graph-based learning, reduced-order modelling, and inverse design~\cite{paul2022graphbased,karapiperis2023prediction, zheng2023unifying}. One can, for example, use the framework to analyze experimental data by correlating extracted geometric characteristics with mechanically tested stress-strain curves and identifying the dominant geometric features of mechanical properties. This provides a quantitative basis for searching pattern layouts that realize the prescribed target spatial distributions of geometric descriptors, or using persistent geometric localization regions as features to identify candidate design constraints in wearables and soft robotic textiles. Another feasible application lies in finite element modeling, where the extracted geometric parameters augment the formulation of elastic energy density for amorphous and heterogeneous materials such as knitted polymer composites~\cite{liu2024compositeI,D5SM01027F,CHEN2026109762,ding2026crack}.

Given the wide applicability of the present work, several notable improvement strategies can be outlined for future research directions. For example, replacing rectangular heatmaps with topological graphs mapped to actual stitch positions and densities will yield more reliable results. Moreover, the randomness of loop relaxation and deformation, alongside complex behaviors like buckling, exceeds the scope of the current model. Therefore, investigations into these special scenarios can provide a more complete understanding of fabric structure.

%%%%%%%%%%%%%%%%%%%%%%%%%%%%%%%%%%%%%%%%%%%%%%%%%%%%%%%%%%%%%%%%%%%%%%%%%%%%%%%%%%%%%%%%%%%%%%%%%%%%%%%%%%%%%%%%%%%%%%%%%%%%%%%%%%%%%%%%%%%%%%%%%
\section*{Methods}
In this section, we describe our proposed methods for quantifying the change in unit fabrics under force, including reconstructing the unit fabric, defining and constructing the unit surface, calculating quantities related to the size and shape of unit fabrics, and measuring and analyzing the spatial and temporal variation of those quantities. 

%%%%%%%%%%%%%%%%%%%%%%%%%%%%%%
\subsection*{Curve quantification}

To develop an analytical workflow applicable to general textile structures, we first developed a curve reconstruction method. This method could be employed to reconstruct a smooth curve when the data points are too sparse (e.g., in our test data, only 16 input points are available for each unit cell). The generated curves could provide an accurate approximation of the yarn spatial structure, which can be used for subsequent fabric structure analysis and computational validation. At the same time, they also serve as smooth boundaries of the unit surface, offering a reliable basis for the subsequent calculation of the surface-based quantities. See SI Section~S1 and Fig.~S1--S2 for more details of the curve quantification methods.

%%%%%%%%%%%%%%%%%%%%%%%%%%%%%%
\subsubsection*{Smooth curve reconstruction by B-spline}
 Given the coordinates of several nodes on the fabric sample, we perform a least-squares B-spline approximation to reconstruct a smooth curve model to describe the geometrical structure of the yarn. This approach relies only on the given data points representing the fabric structure, without any smoothness requirements, making it applicable to general datasets. Additionally, the order of the approximation B-spline curve can be adjusted to a suitable value under specific smoothness requirements, while avoiding overfitting.

%%%%%%%%%%%%%%%%%%%%%%%%%%%%%%
\subsubsection*{Curve curvature and torsion quantification}

To study the geometric properties of the fabric, we first investigate its structure by performing geometrical quantification on a curve level. In particular, the curve curvature measures the degree of bending of a curve at a given point, while the curve torsion describes the extent to which a curve deviates from being planar, that is, the ``twisting'' characteristic of the curve. Therefore, computing the curvature and torsion of the reconstructed curve model will provide a local shape description of the yarn. To achieve this, dense sampling points from the previously constructed curve model are employed to formulate closely located local Frenet frames, which are used to calculate the curvature and torsion at each of the sampling points.

%%%%%%%%%%%%%%%%%%%%%%%%%%%%%%
\subsection*{Surface quantification} 

To accurately extract the surface features of a unit area formed by an irregularly shaped stitch, we explored several surface reconstruction methods. We found that the hybrid technique combining B-spline and ruled surface is the most effective for preserving boundary smoothness with meaningful surface definition. See SI Section~S2, Fig.~S3--S8, and Table S1--S2 for more details of the surface quantification methods with comparisons.

%%%%%%%%%%%%%%%%%%%%%%%%%%%%%%
\subsubsection*{Triangulation}

We first use the given sample points to construct a coarse triangulation in each unit fabric. The resulting values are then used as a reference value to compare with other results to assess the feasibility of each area computation method.

%%%%%%%%%%%%%%%%%%%%%%%%%%%%%%
\subsubsection*{B\'ezier surface}
To achieve a more accurate unit area computation, we attempt to construct a B\'ezier surface from given nodal points. To employ the third-order B\'ezier curve twice, we select 16 points (using either the given nodes or the middle or quadrisection points between two nodes), and divide them into four groups of four points. Then, points in each group serve as control points to generate a B\'ezier curve. After generating four curves, we combine them into a surface control mesh by selecting one point from each curve at a time to get a new set of four points as control points. Using the new control points, a new B\'ezier curve can be constructed. By iterating through the curve points in this manner, it constructs a surface mesh that intersects the original curve, progressively forming the surface. 

Despite the simplicity of the method mentioned above, issues regarding precision are encountered. First, note that the B\'ezier surface generated by this method exhibits significant deviation from the control points, which arises from the nature of B\'ezier curve that it could only fit precisely with the first and last control points while deviating from those intermediate points. If the 16 given control points are used directly, the constructed surface can only coincide with the four corner points, while the twelve intermediate points remain too distant from the surface. Hence, the generated surface fails to precisely fit the target unit surface. Also, we tested the approach of partitioning the 16 points into seven sets of four points each to achieve a better fit. In that case, the constructed surface will exhibit unsmoothness at each shared edge of adjacent quadrilaterals. As a result, the calculated area will also be inaccurate, roughly equivalent to those obtained with direct triangulation, and hence is not ideal.  Moreover, sharp transitions will occur at the shared edges of adjacent quadrilaterals, causing significant errors in surface curvature calculations. Finally, this two-parameter mesh surface construction method may result in distorted or self-intersecting surfaces, since it computes the mesh grid directly according to the input order, irrespective of any spatial characteristics of the nodes.

%%%%%%%%%%%%%%%%%%%%%%%%%%%%%%
\subsubsection*{Improved B\'ezier surface}

Given the limitations of the methods above, we aim to improve the construction methods for B\'ezier surfaces. Instead of using the original nodes directly as control points, we select new control points before forming the B\'ezier curve. We first perform parameterization to restrict the range of the control points, and then compute the bicubic form of the basis function matrix for the B\'ezier curve. Using the bicubic basis function matrix and the coordinates of the original nodes, we solve for the least-squares solution for the new control points using the pseudo-inverse matrix. As a consequence, the new B\'ezier curve will have the smallest deviation from the given nodes. 

Through the aforementioned improvements, we have resolved the issue of excessive deviation between the surface and control points. Simultaneously, by selecting the new control points and imposing constraints on the spatial order of control points, specifically requiring that the $x$-coordinates of control points in each row and the $y$-coordinates in each column be arranged in ascending order, we have addressed surface distortion and self-intersection concerns. Nevertheless, this approach still possesses certain drawbacks. First, discretizing the parameter $u$ with a fixed step size may result in insufficient precision or unsatisfying computation speed, such as failing to achieve a finer mesh in regions of high curvature, or low efficiency if better accuracy is required. Therefore, adaptive step size adjustment is needed for better performance. Second, the point cloud generated by the above approach will extend beyond the unit surface region as a result of the mesh generation step. Specifically, in the method, we first form the $u,v$ grid based on $x_{\min} < x_{\text{point}} < x_{\max}$ and $y_{\min} < y_{\text{point}} < y_{\max}$, ensuring the generated surface projects as a rectangle onto the $x$-$y$ plane. Consequently, we need to add extra boundaries to each B\'ezier curve to find the surface with the fabric being its boundary curve.

To calculate the surface area of the unit surface depicted by the point cloud generated as above, the Delaunay triangulation can be considered. The points are projected onto a certain plane (e.g., the $x$-$y$ plane) before doing the Delaunay triangulation. Consequently, this method requires dividing the surface into partitions whose projection onto the $x$-$y$ plane are all convex hulls, and we also need to find suitable projection angles to ensure the corresponding surface returned is the correct one we desire. However, the multilayer structure and the continuous rotation of the unit fabric make the partition and projection steps complicated, since it is hard to find suitable projection angles. Hence, a more sophisticated algorithm, including adaptive step size, restricted B\'ezier curves, and a partition and projection scheme for the generated point cloud, is needed if we want to proceed with this surface reconstruction method. In consideration of the above shortcomings, we aim to explore more direct and efficient methods for surface construction.

%%%%%%%%%%%%%%%%%%%%%%%%%%%%%%
\subsubsection*{Ruled surface}
As we have reconstructed the boundary curve using the B-spline curve during the curve smoothing procedure, we may utilize the previous result to gain a consistent definition of the unit surface. At the same time, inspired by the idea of fitting surfaces with multiple curves from the aforementioned B\'ezier surfaces, we hope to start from the set of boundary points and obtain a unit surface through drawing dense lines.

In earlier computation, we attained the coordinates of a set of points on the boundary curve. Using this known result, the cubic spline that depicts the yarn itself will be defined as the directrix of the ruled surface, whereas every two points will be connected in a head-to-tail manner as a pair to form segments as generators. Having drawn the surface using the above method, we then take $n_l$ equally spaced points on each generator. In this way, we obtain a mesh of sampling points on the ruled surface with the boundary curve being the given B-spline curve. A finer mesh could be obtained with a larger number of sampling points on the boundary curve and generators, and consequently, surface reconstruction with higher precision can be achieved.

%%%%%%%%%%%%%%%%%%%%%%%%%%%%%%
\subsubsection*{Unit area calculation by the B-spline boundary curve}

Following the aforementioned convention of constructing the unit surface as the ruled surface, whose boundary curve is the reconstructed B-spline curve, we form the triangulation using points on the boundary curve. Using more sampling points, we could decompose the ruled surface into a triangular mesh similar to the direct triangulation with 16 nodes, while gaining a more refined triangulation and thus a more accurate result. Calculating the area of each triangle and summing them up will give the approximate unit area defined by the ruled surface. This method ensures that the triangular mesh has no overlaps and fully covers the unit area bounded by the B-spline curve, consistent with the previous definition of the ruled surface.

%%%%%%%%%%%%%%%%%%%%%%%%%%%%%%
\subsubsection*{Surface curvature quantification}

Through the ruled surface reconstruction, we obtain a surface mesh. With the vertex coordinates of the surface mesh, the partial derivatives can be calculated using the finite difference method. Then, the coefficients of the first and second fundamental forms of the surface at discrete points can be calculated accordingly, from which we can obtain the surface Gaussian curvature and surface mean curvature.

%%%%%%%%%%%%%%%%%%%%%%%%%%%%%%
\subsection*{Volume quantification} 

In this section, we aim to calculate the volume and aspect ratio of a unit fabric, which will provide important information about the shape and size of the unit fabric. However, neither curves nor surfaces have a definition of volume. Therefore, we considered different definitions and examined their accuracy and adaptability to complex structures and varying spatial positions. See SI Section~S3, Fig.~S9 and Table~S3 for more details of the volume quantification methods with comparisons.
 
\subsubsection*{Axis-aligned bounding box}
We first consider the axis-aligned bounding box approach, in which a box with edges parallel to the coordinate axes is used to enclose the given unit fabric. Then, we can use the bounding box to compute the volume of a unit fabric, and the aspect ratio is given by the length-to-width ratio of the bounding box. Correspondingly, the centroid of the bounding box is used as the position of the unit fabric for subsequent analysis.
 
Although the above method is conceptually and computationally simple, it has a notable limitation. Specifically, as the bounding box is always aligned with the coordinate axes, it cannot fit tightly to irregularly shaped objects. Note that the unit fabric has a sheet-like structure sensitive to rotation, so the bounding box will contain a lot of redundant space in some states. Moreover, the cylindrical fabric considered in our work has unit fabrics with different angles, and hence this method may lead to significant errors in volume calculation.

\subsubsection*{Minimal bounding box}
Another approach is to consider the minimal bounding box, which does not depend on the alignment with the coordinate axes. Specifically, the minimal bounding box for a set of points is the smallest orientation-free rectangular box that completely encloses all points in the 3D space. It can be rotated to align with the principal directions of the point set, thereby providing a tighter fit. 

After obtaining the minimal bounding box, we can calculate its volume using its dimensions. Also, using the centroid of the minimal bounding box, we can define the position of the unit fabric. As for the aspect ratio, recall that we used the length-to-width ratio of the bounding box in the previous approach. However, since the minimal bounding box is not necessarily aligned with the coordinate axes, there is an ambiguity in the definition of ``length'' and ``width''. To resolve this issue, we calculate the aspect ratio as the ratio of the maximum distance of points in the point cloud along the $x$ and $y$ directions.

Although this approach can effectively reduce the error brought by the rotation of the fabric pieces, it cannot take into account the complex structure of the units and may still contain a lot of redundant space, thereby leading to inaccuracies in the volume calculation.

\subsubsection*{Convex hull}
Noticing the drawbacks of the two above-mentioned approaches, we further consider using the convex hull to compute the volume. In three-dimensional space, the convex hull of a point cloud refers to the smallest convex polyhedron that contains all the points in the set. This means that the line segment connecting any two points in the point cloud is a subset of the convex hull, and its vertices form several convex polygonal faces. Using this method, we can reduce redundant space and make the calculation more adaptable to irregular shapes.
    
%%%%%%%%%%%%%%%%%%%%%%%%%%%%%%
\subsection*{Quantifying spatial and temporal variations}

After quantifying the geometric features of the unit fabrics using the above curve, surface, and volume-based methods, we aim to analyze the mechanical properties of each fabric by comparing how these quantities change in each unit under external forces. To achieve this, we measure both temporal and spatial changes of the above quantities, and create heatmaps and line graphs for illustration.

More specifically, we have a consistent temporal framework to formalize the calculation of temporal variation and align the horizontal axis of line graphs of averaged spatial variation. The following clarifies the mapping between our discrete equilibrium sampling and the temporal scale adopted throughout this work.

The yarn structural dataset employed in this work comprises a series of discrete equilibrium states acquired under two distinct loading configurations:
\begin{itemize}
    \item ‌\textbf{Uniaxial in-plane tension‌:} For planar knitted fabric specimens, equilibrium states are captured across the full strain range from 0\% to 100\% at 20\% strain increments;
    \item ‌\textbf{Bending loading‌:} For the cylindrical fabric specimen, equilibrium states are reconstructed across the full bending angle range from $0^{\circ}$ to $60^{\circ}$ at $1^{\circ}$ increments.
\end{itemize}

Accordingly, the evolution of geometric descriptors investigated here is parameterized along the ‌quasi-static loading path‌, and the temporal scale in this context specifically corresponds to the ‌loading history scale‌ defined by these incrementally increasing state points.

\subsubsection*{Spatial variation}

We compute the course-wise and wale-wise spatial derivatives using the finite difference method. The course-wise and wale-wise distances between each pair of adjacent elements are first calculated (denoted as $\Delta x_{x}$ and $\Delta x_{y}$). Next, by performing a first-order difference operation on a two-dimensional matrix along the first and second dimensions, the differences of each quantity between adjacent elements are computed (denoted as $\Delta Q_x$ and $\Delta Q_y$). The difference results, along with the spatial step size, essentially provide a discrete approximation of the partial derivatives, corresponding to the numerical implementation of $\frac{\partial Q}{\partial x}  \approx \frac{\Delta Q}{\Delta x}  $, where $Q$ can be substituted by quantities computed previously (e.g., fabric curvature, area, aspect ratio, etc.).

After obtaining the spatial variation of each unit, we get two matrices at each time point, the units of which represent the course-wise and wale-wise spatial variation. We then summarize the result by taking the average of the absolute value of spatial variation over all the units and drawing two line graphs to demonstrate the trend. These line graphs provide us with an overview of the spatial difference of certain geometric quantities of the fabric at each time point, as well as the trend in their variation.

\subsubsection*{Temporal variation}

We can further compute the temporal variation of each quantity by taking the difference between the initial frames and the intermediate frames ($ \Delta Q _t = Q_t - Q_0$). By dividing the difference by the initial amount, we get a percentage change relative to the initial amount ($P_t=\frac{\Delta Q_t}{Q_0}$). This method is dimension-independent and hence enables the comparison between data with different units. By eliminating the incomparability caused by different dimensions, data with different dimensions could be involved in model calculation at the same scale. 

The above calculation gives us matrices at each time point showing the percentage variation of each unit with respect to the initial amount. We can then plot heatmaps to illustrate the result, and the combination of these figures will show the quantity change over a time period. By comparing the dynamic behaviors in different zones of each sample and the heatmaps of different knitting patterns, we can analyze their mechanical properties accordingly.

\subsubsection*{Interpretation of hot spots}
Throughout this paper, the term ``hot spot'' denotes a region of persistently large absolute geometric variation for a chosen descriptor. Unless otherwise stated, this terminology refers to geometric localization and does not by itself imply a direct measurement of stress, energy density, or damage. However, our framework provides a quantitative language for nonlinear deformation in knits and establishes a geometry-based representation that can be coupled to constitutive models, in order to provide stress-, energy-, and damage- related mechanical interpretations.

\bibliographystyle{ieeetr}
\bibliography{reference}

\begin{thebibliography}{10}

\bibitem{bertoldi2017flexible}
K.~Bertoldi, V.~Vitelli, J.~Christensen, and M.~Van~Hecke, ``Flexible mechanical metamaterials,'' {\em Nat. Rev. Mater.}, vol.~2, no.~11, pp.~1--11, 2017.

\bibitem{Moestopo2023}
W.~P. Moestopo, S.~Shaker, W.~Deng, and J.~R. Greer, ``Knots are not for naught: Design, properties, and topology of hierarchical intertwined microarchitected materials,'' {\em Sci. Adv.}, vol.~9, no.~10, p.~eade6725, 2023.

\bibitem{surjadi2025double}
J.~U. Surjadi, B.~F. Aymon, M.~Carton, and C.~M. Portela, ``Double-network-inspired mechanical metamaterials,'' {\em Nat. Mater.}, vol.~24, pp.~945--954, 2025.

\bibitem{pescialli2025topology}
E.~Pescialli, A.~{Munoz Lopez}, K.~Karapiperis, and D.~M. Kochmann, ``Topology-informed design of intertwined architected materials: Unifying woven, knotted, and closed-chain networks,'' {\em Mater. Des.}, vol.~260, p.~114974, 2025.

\bibitem{choi2019programming}
G.~P.~T. Choi, L.~H. Dudte, and L.~Mahadevan, ``Programming shape using kirigami tessellations,'' {\em Nat. Mater.}, vol.~18, no.~9, pp.~999--1004, 2019.

\bibitem{choi2021compact}
G.~P.~T. Choi, L.~H. Dudte, and L.~Mahadevan, ``Compact reconfigurable kirigami,'' {\em Phys. Rev. Research}, vol.~3, p.~043030, Oct 2021.

\bibitem{hong2022boundary}
Y.~Hong, Y.~Chi, S.~Wu, Y.~Li, Y.~Zhu, and J.~Yin, ``Boundary curvature guided programmable shape-morphing kirigami sheets,'' {\em Nat. Commun.}, vol.~13, no.~1, p.~530, 2022.

\bibitem{dudek2025shape}
K.~K. Dudek, M.~Kadic, C.~Coulais, and K.~Bertoldi, ``Shape-morphing metamaterials,'' {\em Nat. Rev. Mater.}, vol.~10, pp.~783--798, 2025.

\bibitem{el2021digital}
C.~El~Helou, P.~R. Buskohl, C.~E. Tabor, and R.~L. Harne, ``Digital logic gates in soft, conductive mechanical metamaterials,'' {\em Nat. Commun.}, vol.~12, no.~1, p.~1633, 2021.

\bibitem{mohammadi2021flexible}
A.~Mohammadi, Y.~Tan, P.~Choong, and D.~Oetomo, ``Flexible mechanical metamaterials enabling soft tactile sensors with multiple sensitivities at multiple force sensing ranges,'' {\em Sci. Rep.}, vol.~11, no.~1, p.~24125, 2021.

\bibitem{jiao2023mechanical}
P.~Jiao, J.~Mueller, J.~R. Raney, X.~Zheng, and A.~H. Alavi, ``Mechanical metamaterials and beyond,'' {\em Nat. Commun.}, vol.~14, no.~1, p.~6004, 2023.

\bibitem{luo2022digital}
Y.~Luo, K.~Wu, A.~Spielberg, M.~Foshey, D.~Rus, T.~Palacios, and W.~Matusik, ``Digital fabrication of pneumatic actuators with integrated sensing by machine knitting,'' {\em Proceedings of the 2022 CHI Conference on Human Factors in Computing Systems}, pp.~1--13, 2022.

\bibitem{sanchez20233d}
V.~Sanchez, K.~Mahadevan, G.~Ohlson, M.~A. Graule, M.~C. Yuen, C.~B. Teeple, J.~C. Weaver, J.~McCann, K.~Bertoldi, and R.~J. Wood, ``{3D} knitting for pneumatic soft robotics,'' {\em Adv. Funct. Mater.}, vol.~33, no.~26, p.~2212541, 2023.

\bibitem{wang2024soft}
M.~Wang, Y.~Zhou, and R.~Stewart, ``Soft wearable robotics: Innovative knitting-integrated approaches for pneumatic actuators design,'' in {\em Companion Publication of the 2024 ACM Designing Interactive Systems Conference}, pp.~234--238, 2024.

\bibitem{luo2021}
Y.~Luo, Y.~Li, P.~Sharma, W.~Shou, K.~Wu, M.~Foshey, B.~Li, T.~Palacios, A.~Torralba, and W.~Matusik, ``Learning human–environment interactions using conformal tactile textiles,'' {\em Nat. Electron.}, vol.~4, pp.~193--201, 2021.

\bibitem{wan2024all}
X.~Wan, Y.~Shen, T.~Luo, M.~Xu, H.~Cong, C.~Chen, G.~Jiang, and H.~He, ``All-textile piezoelectric nanogenerator based on {3D} knitted fabric electrode for wearable applications,'' {\em ACS Sensors}, vol.~9, no.~6, pp.~2989--2998, 2024.
\newblock PMID: 38771707.

\bibitem{kou2024wearable}
Z.~Kou, C.~Zhang, B.~Yu, H.~Chen, Z.~Liu, and W.~Lu, ``Wearable all-fabric hybrid energy harvester to simultaneously harvest radiofrequency and triboelectric energy,'' {\em Adv. Sci.}, vol.~11, no.~17, p.~2309050, 2024.

\bibitem{zhou2024highly}
Y.~Zhou, Y.~Sun, Y.~Li, C.~Shen, Z.~Lou, X.~Min, and R.~Stewart, ``A highly durable and {UV}-resistant graphene-based knitted textile sensing sleeve for human joint angle monitoring and gesture differentiation,'' {\em Adv. Intell. Syst.}, vol.~6, no.~10, p.~2400124, 2024.

\bibitem{han2017blooming}
M.-W. Han and S.-H. Ahn, ``Blooming knit flowers: loop-linked soft morphing structures for soft robotics,'' {\em Adv. Mater.}, vol.~29, no.~13, p.~1606580, 2017.

\bibitem{ali2017knitting}
A.~Maziz, A.~Concas, A.~Khaldi, J.~Stålhand, N.-K. Persson, and E.~Jager, ``Knitting and weaving artificial muscles,'' {\em Sci. Adv.}, vol.~3, no.~1, p.~e1600327, 2017.

\bibitem{FU2022137241}
C.~Fu, K.~Wang, W.~Tang, A.~Nilghaz, C.~Hurren, X.~Wang, W.~Xu, B.~Su, and Z.~Xia, ``Multi-sensorized pneumatic artificial muscle yarns,'' {\em Chem. Eng. J.}, vol.~446, p.~137241, 2022.

\bibitem{paras2020geometric}
P.~Wadekar, P.~Goel, C.~Amanatides, G.~Dion, R.~D. Kamien, and D.~E. Breen, ``Geometric modeling of knitted fabrics using helicoid scaffolds,'' {\em J. Eng. Fibers Fabr.}, vol.~15, p.~1558925020913871, 2020.

\bibitem{kapllani2021topoknit}
L.~Kapllani, C.~Amanatides, G.~Dion, V.~Shapiro, and D.~E. Breen, ``{TopoKnit}: A process-oriented representation for modeling the topology of yarns in weft-knitted textiles,'' {\em Graph. Models}, vol.~118, p.~101114, 2021.

\bibitem{kapllani2022loop}
L.~Kapllani, C.~Amanatides, G.~Dion, and D.~E. Breen, ``Loop order analysis of weft-knitted textiles,'' {\em Textiles}, vol.~2, no.~2, pp.~275--295, 2022.

\bibitem{rout2022}
S.~K. Rout, M.~R. Bisram, and J.~Cao, ``Methods for numerical simulation of knit based morphable structures: knitmorphs,'' {\em Scientific Reports}, vol.~12, 12 2022.

\bibitem{niu2025geometric}
L.~Niu, G.~Dion, and R.~D. Kamien, ``Geometric modeling of knitted fabrics,'' {\em Proc. Natl. Acad. Sci. U. S. A.}, vol.~122, no.~7, p.~e2416536122, 2025.

\bibitem{bergou2008discrete}
M.~Bergou, M.~Wardetzky, S.~Robinson, B.~Audoly, and E.~Grinspun, ``Discrete elastic rods,'' in {\em ACM SIGGRAPH 2008 Papers}, SIGGRAPH '08, (New York, NY, USA), Association for Computing Machinery, 2008.

\bibitem{Kaldor2010}
J.~M. Kaldor, D.~L. James, and S.~Marschner, ``Efficient yarn-based cloth with adaptive contact linearization,'' {\em ACM Trans. Graph.}, vol.~29, pp.~1--10, 2010.

\bibitem{ding2024unravelling}
X.~Ding, V.~Sanchez, K.~Bertoldi, and C.~H. Rycroft, ``Unravelling the mechanics of knitted fabrics through hierarchical geometric representation,'' {\em Proc. R. Soc. A}, vol.~480, no.~2295, p.~20230753, 2024.

\bibitem{Vic2D}
{Vic2D Development Team}, ``{Vic2D}: Digital image correlation software,'' 2023.
\newblock Accessed: 2025-11-13.

\bibitem{DICengine}
{DICengine Development Team}, ``{DICengine}: Digital image correlation engine,'' 2024.
\newblock Accessed: 2025-11-13.

\bibitem{ROUX20081253}
S.~Roux, F.~Hild, P.~Viot, and D.~Bernard, ``Three-dimensional image correlation from {X}-ray computed tomography of solid foam,'' {\em Compos. - A: Appl. Sci. Manuf.}, vol.~39, no.~8, pp.~1253--1265, 2008.
\newblock Full-field Measurements in Composites Testing and Analysis.

\bibitem{paul2022graphbased}
P.~P. Meyer, C.~Bonatti, T.~Tancogne-Dejean, and D.~Mohr, ``Graph-based metamaterials: Deep learning of structure-property relations,'' {\em Mater. Des.}, vol.~223, p.~111175, 2022.

\bibitem{karapiperis2023prediction}
K.~Karapiperis and D.~M. Kochmann, ``Prediction and control of fracture paths in disordered architected materials using graph neural networks,'' {\em Commun. Eng.}, vol.~2, no.~1, p.~32, 2023.

\bibitem{zheng2023unifying}
L.~Zheng, K.~Karapiperis, S.~Kumar, and D.~M. Kochmann, ``Unifying the design space and optimizing linear and nonlinear truss metamaterials by generative modeling,'' {\em Nat. Commun.}, vol.~14, no.~1, p.~7563, 2023.

\bibitem{huang2024bending}
Y.~Huang, H.~Ren, Y.~Liu, W.~Xu, and W.~Zhao, ``Bending shape memory properties and multi-scale viscoelastic behaviors of knitted-fabric reinforced polymer composites,'' {\em Compos. Sci. Technol.}, vol.~256, p.~110747, 2024.

\bibitem{yang2026shape}
W.~Yang, Y.~Liu, Y.~Ouyang, W.~Xu, W.~Zhao, and H.~Ren, ``Shape memory performance and mechanical properties of natural fiber-reinforced polymer composites enhanced via fabrication strategies,'' {\em Ind. Crops Prod.}, vol.~239, p.~122546, 2026.

\bibitem{bai20263d}
T.~Bai, X.~Qin, Z.~Wu, J.~Su, L.~Li, H.~Yao, Z.~Niu, C.~Yue, D.~Wang, Y.~Yue, {\em et~al.}, ``{3D}-printed resilient biomass-based covered stents for tracheal implants,'' {\em Adv. Compos. Hybrid Mater.}, 2026.

\bibitem{patil2020topological}
V.~P. Patil, J.~D. Sandt, M.~Kolle, and J.~Dunkel, ``Topological mechanics of knots and tangles,'' {\em Science}, vol.~367, no.~6473, pp.~71--75, 2020.

\bibitem{TONG2026102430}
D.~Tong, A.~Choi, J.~Wang, W.~Huang, Z.~Chen, J.~Li, X.~Huang, M.~Liu, H.~Gao, and K.~J. Hsia, ``Discrete differential geometry for simulating nonlinear behaviors of flexible systems: A survey,'' {\em Extreme Mech. Lett.}, vol.~82, p.~102430, 2026.

\bibitem{yao2025experimental}
H.~Yao, T.~Bai, Z.~Han, Z.~Niu, J.~Su, J.~Yan, J.~Lu, D.~Wang, W.~Zhao, G.~Han, {\em et~al.}, ``Experimental and numerical investigation on low-velocity impact of plain-woven bamboo fiber reinforced epoxy resin composites based on multiscale modeling,'' {\em Compos. B: Eng.}, vol.~306, p.~112779, 2025.

\bibitem{RADI2023104658}
K.~Radi, F.~Allamand, and D.~M. Kochmann, ``Deformation tracking of truss lattices under dynamic loading based on digital image correlation,'' {\em Mech. Mater.}, vol.~183, p.~104658, 2023.

\bibitem{liu2024compositeI}
F.~Liu, X.~Chen, Z.~Suo, and J.~Tang, ``Composite of knitted fabric and soft matrix. {I}. crack growth in the course direction,'' {\em Soft Matter}, vol.~20, p.~9614, 2024.

\bibitem{D5SM01027F}
F.~Liu, W.~Zhang, S.~Zheng, X.~Chen, and J.~Tang, ``Composite of knitted fabric and soft matrix. {II}. laddering,'' {\em Soft Matter}, vol.~22, pp.~1230--1239, 2026.

\bibitem{CHEN2026109762}
X.~Chen, Y.~Lu, W.~Zhang, S.~Zheng, F.~Liu, and J.~Tang, ``Stretchable composites with high initiation fatigue threshold comparable to soft tissues,'' {\em Int. J. Fatigue}, vol.~212, p.~109762, 2026.

\bibitem{ding2026crack}
X.~Ding, Y.~Niu, Y.~Zhao, H.~Chen, C.~Bae, A.~T. Nguyen, H.~D. Espinosa, Z.~P. Bažant, and J.~Cao, ``Crack-parallel tension effects on fracture of soft knitted polymer composites deduced from gap tests via microplane triads,'' {\em Preprint}, 06 2026.
\newblock Available at: https://www.mechanicsarxiv.org/index.php/engineering/preprint/view/100.

\bibitem{frenetrobust}
B.~Friedrich, ``frenet\_robust, {MATLAB} {C}entral {F}ile {E}xchange,'' 2014.
\newblock \url{https://www.mathworks.com/matlabcentral/fileexchange/47885-frenet_robust-zip}, accessed on November 18, 2025.

\bibitem{Bezier}
S.~Mohanty, ``Bezier {S}urface, {MATLAB} {C}entral {F}ile {E}xchange,'' 2015.
\newblock \url{https://www.mathworks.com/matlabcentral/fileexchange/66678-bezier-surface}, accessed on November 18, 2025.

\bibitem{minboundbox}
J.~Korsawe, ``Minimal {B}ounding {B}ox, {MATLAB} {C}entral {F}ile {E}xchange,'' 2015.
\newblock \url{https://www.mathworks.com/matlabcentral/fileexchange/18264-minimal-bounding-box}, accessed on November 18, 2025.

\end{thebibliography}

\clearpage

%%%%%%%%%%%%%%%%%%%%%%%%%%%%%%%%%%%%%%
%  SI 

\graphicspath{{./FIGURE/SI/}}

\centerline{\Large\textbf{Supplementary Information}}
\appendix
\renewcommand\thefigure{S\arabic{figure}}    
\setcounter{figure}{0}
\renewcommand\thetable{S\arabic{table}}    
\setcounter{table}{0}
\renewcommand{\thesection}{S\arabic{section}}

\section{Curve quantification}

Here we present the mathematical details of our curve quantification methods. Using the B-spline structure, smooth fabric curves could be reconstructed, and subsequent curve curvature and torsion quantification steps could be performed based on the resulting model. 

\subsection{Smooth curve reconstruction by B-spline}
 We are initially given $x$, $y$, and $z$ coordinates of 16 nodes in each unit of fabric, from which we aim to construct a smooth curve depicting the yarn itself. Since the coordinates in three dimensions are stored separately in three block matrices of size $16 \times 1$, we perform least-squares B-spline approximation to calculate the curve in each of the three dimensions independently. 

 Take data in $x$-coordinates as an example: Consider the B-spline function parametrized by $u \in [0,1]$. We first generate 16 uniformly distributed parameter values $u$, ranging from 0 to 1, and parameterize the control points to the $[0,1]$ interval. Meanwhile, fourth-order B-spline basis functions are created, and a B-spline curve that minimizes the total squared deviation from all control points is obtained using least-squares fitting. 

Mathematically, the curve is expressed as:
\[
C(u) = \sum_{i=0}^{m} P_i N_{i,p}(u),
\]
where $P_i$ are the control points, $N_{i,p}(u)$ are the $p$-degree B-spline basis functions ($p=4$ in our case) with
    \[
    N_{i,0}(u) = 
    \begin{cases} 
    1 & \text{if } u_i \leq u < u_{i+1}, \\
    0 & \text{otherwise},
    \end{cases}
    \]
    \[
    N_{i,p}(u) = \frac{u - u_i}{u_{i+p} - u_i} N_{i,p-1}(u) + \frac{u_{i+p+1} - u}{u_{i+p+1} - u_{i+1}} N_{i+1,p-1}(u),
    \]
    and $u$ is the parameter defined on the interval $[0,1]$.

By minimizing the error between the curve and the data points, the approximation B-spline could be obtained. Specifically, we need to solve for control points $P_i$ that minimize the following objective function:
\[
\min_{P_0, P_1, \ldots, P_m} \sum_{k=0}^{n} \| C(u_k) - D_k \|^2,
\]
where $u_k$ are the parameterized parameter values and $D_k$ is the $k$-th data point.

The above problem can be transformed into a linear system of equations, represented as:
\[
\bm{A} \bm{P} = \bm{D},
\]
where $\bm{A}$ is the basis function matrix with elements $A_{k,i} = N_{i,p}(u_k)$, $\bm{P}$ is the control point vector $\bm{P} = [P_0, P_1, \ldots, P_m]^T$, and $D$ is the data point vector $\bm{D} = [D_0, D_1, \ldots, D_n]^T$. The solution is obtained through the least squares method:
\[
\bm{P} = (\bm{A}^T \bm{A})^{-1} \bm{A}^T \bm{D}.
\]
After obtaining the B-spline structure, $n$ equally spaced new parameter values $u$ are generated ($u \in [0,1]$), after which $n$ $x$-coordinate values corresponding to the $n$ sampling parameter points can be calculated using the spline structure. By adjusting n, the sampling density of the output curve could be controlled to balance speed and accuracy. The resulting data are stored in an $n \times 1$ matrix and output as the $x$-coordinates of the $n$ new sampling points. The $y$ and $z$ coordinates are processed in the same way, and the three-dimensional coordinates of the $n$ new sampling points could then be obtained.

\begin{figure}[t]
    \centering 
    \includegraphics[width=0.6\linewidth]{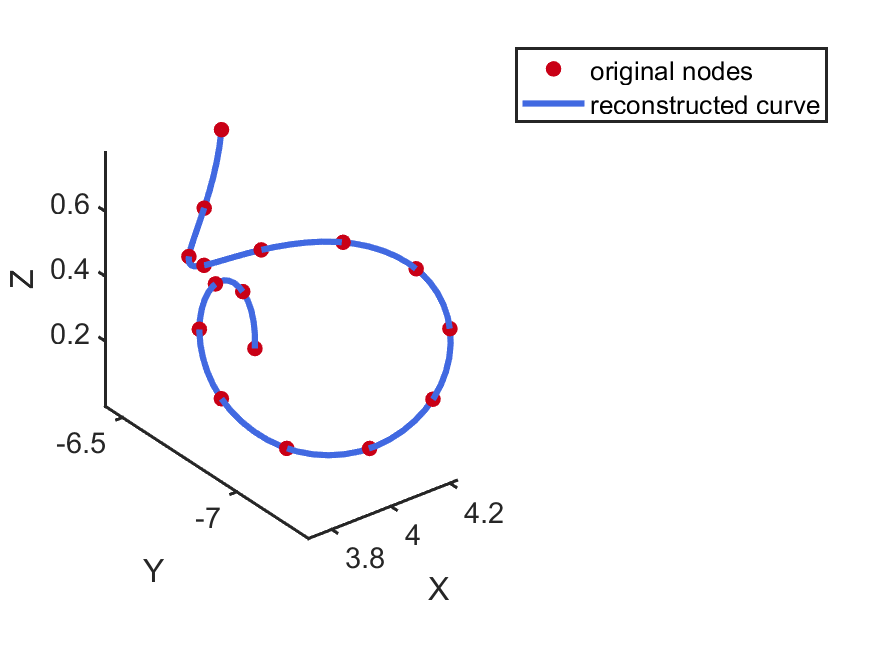}
    \caption{\textbf{Curve reconstruction using B-spline approximation.} Here, the red points indicate the original vertices, and the blue curve represents the reconstructed smooth curve. }
    \label{curverecon}
    \end{figure}

Through the above process, we obtain a B-spline curve consisting of $n$ sampling points that approximates the control points while maintaining smoothness and continuity properties (see Fig.~\ref{curverecon} for an example).

\subsection{Curve curvature and torsion quantification}
    
Having obtained the reconstructed boundary curve, we can analyze the local geometric structure of the unit fabric by computing its curvature and torsion, which are essential components of a complete quantification framework. Specifically, curvature measures the degree of bending of a curve at a given point, while torsion describes the extent to which a curve deviates from being planar, reflecting the ``twisting" characteristic of the curve.

Mathematically, the curvature of a parametric space curve $\mathbf{r}: I \to \mathbb{R}^3$ with $\mathbf{r}(u)= (x(u),y(u),z(u))$, where $I$ is a parameter interval, is given by:
    \[
    \kappa(u) = \frac{\|\mathbf{r}'(u) \times \mathbf{r}''(u)\|}{\|\mathbf{r}'(u)\|^3},
    \]
where $\mathbf{r}'(u)$ and $\mathbf{r}''(u)$ are the first and second derivatives of the curve, respectively. The torsion of the curve is given by 
    \[
    \tau(u) = \frac {\left({\mathbf{r}'(u)\times \mathbf{r}''(u)}\right)\cdot \mathbf{r}'''(u)}{\left\|{\mathbf{r}'(u)\times \mathbf{r}''(u)}\right\|^{2}}.
    \]

If we further consider the arc length parameterization, i.e., a re-parameterization of the curve by its length: $\mathbf{r}: [0,L] \to \mathbb{R}^3$ where $L$ is the total length of the curve, then we have $\|\mathbf{r}'(s)\|=1$ for all $s \in [0,L]$ and hence
\[
\kappa(s) = \|\mathbf{r}'(s) \times \mathbf{r}''(s)\| = \|\mathbf{T}'(s)\|,
\]
where $\mathbf{T}(s) = \frac{\mathbf{r}'(s)}{\|\mathbf{r}'(s)\|} = \mathbf{r}'(s)$ is the tangent vector. Analogously, the torsion can be simplified as
\[
\tau(s) = -\mathbf{B}'(s) \cdot \mathbf{N}(s),
\]
where $\mathbf{N}(s) = \frac{\mathbf{T}'(s)}{\|\mathbf{T}'(s)\|}$ is the normal vector and $\mathbf{B}(s) = \mathbf{T}(s) \times \mathbf{N}(s)$ is the binormal vector. 

\begin{figure}[t]
    \centering 
    \includegraphics[width=0.6\linewidth]{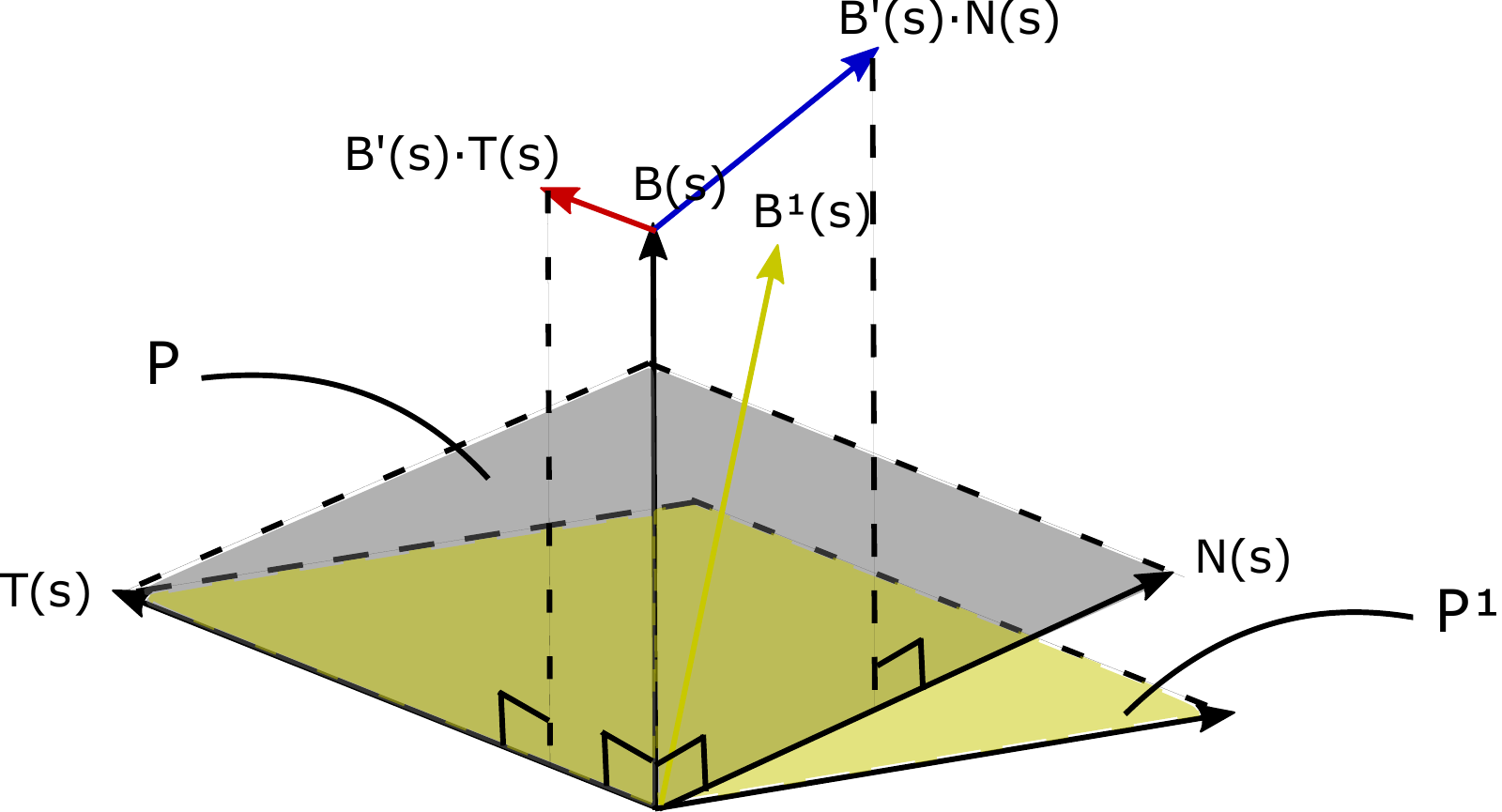}
    \caption{\textbf{Geometric meaning of the torsion of a curve.} In the figure, $\mathbf{T}(s)$, $\mathbf{N}(s)$, $\mathbf{B}(s)$ represent the tangent vector, normal vector, and binormal vector respectively. $\mathbf{P}$ represents the osculating plane spanned by the normal vector and the tangent vector, which is perpendicular to the binormal vector. $\mathbf{B}'(s) \cdot \mathbf{N}(s)$ represents the projection of the variation rate of the binormal vector to the normal vector. $\mathbf{B^{1}}(s)$ represents the variational tendency of the binormal vector in the normal direction, and $\mathbf{P^{1}}$ represents the rotational tendency of the osculating plane about the tangent vector. Here, we only consider the variation of the binormal vector in the direction of the normal vector, and the corresponding rotation of the osculating plane about the tangent vector, to visualize the quantity described by torsion. There is also possibly a variation of the binormal vector in the direction of the tangent vector (as shown in the figure, denoted by $\mathbf{B}'(s) \cdot \mathbf{T}(s)$), and hence the rotation of the osculating plane about the normal vector.}
    \label{torsion}
    \end{figure}

In Fig.~\ref{torsion}, we further explain the geometric meaning of the torsion. Note that the osculating plane of a curve is the plane spanned by the tangent vector $\mathbf{T}(s)$ and the normal vector $\mathbf{N}(s)$, while the binormal vector $\mathbf{B}(s)$ is a unit vector perpendicular to the osculating plane, whose direction directly determines the orientation of the osculating plane. Hence, the variation rate $\mathbf{B}'(s)$ of the binormal vector reflects the rotation of the osculating plane. Note that the formula for torsion quantifies the projection of the variation rate of the binormal vector onto the normal vector. As such, the magnitude of torsion directly reflects the rotational rate of the osculating plane about the tangent vector, serving as a metric for describing the ``twisting" degree of the curve. The greater the torsion, the more pronounced the curve's deviation from planarity, and the faster the osculating plane rotates.

In practice, the curvature and torsion can be computed by substituting the vertex coordinates of the reconstructed B-spline curve into the \texttt{frenet\_robust} function in MATLAB~\cite{frenetrobust}. This function performs local geometric analysis on a 3D space curve using a sliding window approach, calculating the curvature and torsion at each sampling point. It estimates the tangent vector through linear regression, fits an optimal fusion plane via singular value decomposition, and applies the Taubin circle fitting algorithm to determine local curvature. Moreover, the torsion is computed as the rate of change of the binormal vector, with Bayesian prior optimization ensuring continuity between adjacent windows.

\section{Surface quantification} 

To capture the surface-based features of the fabric structures, we considered different approaches for surface reconstruction and the computation of the area and curvature quantities. 

\subsection{Triangulation}

As mentioned in the main text, a coarse triangulation in each unit stitch is performed to get a reference value for comparing with the result obtained in subsequent area computation methods.
 
\begin{figure}[t]
    \centering 
    \includegraphics[width=0.8\linewidth]{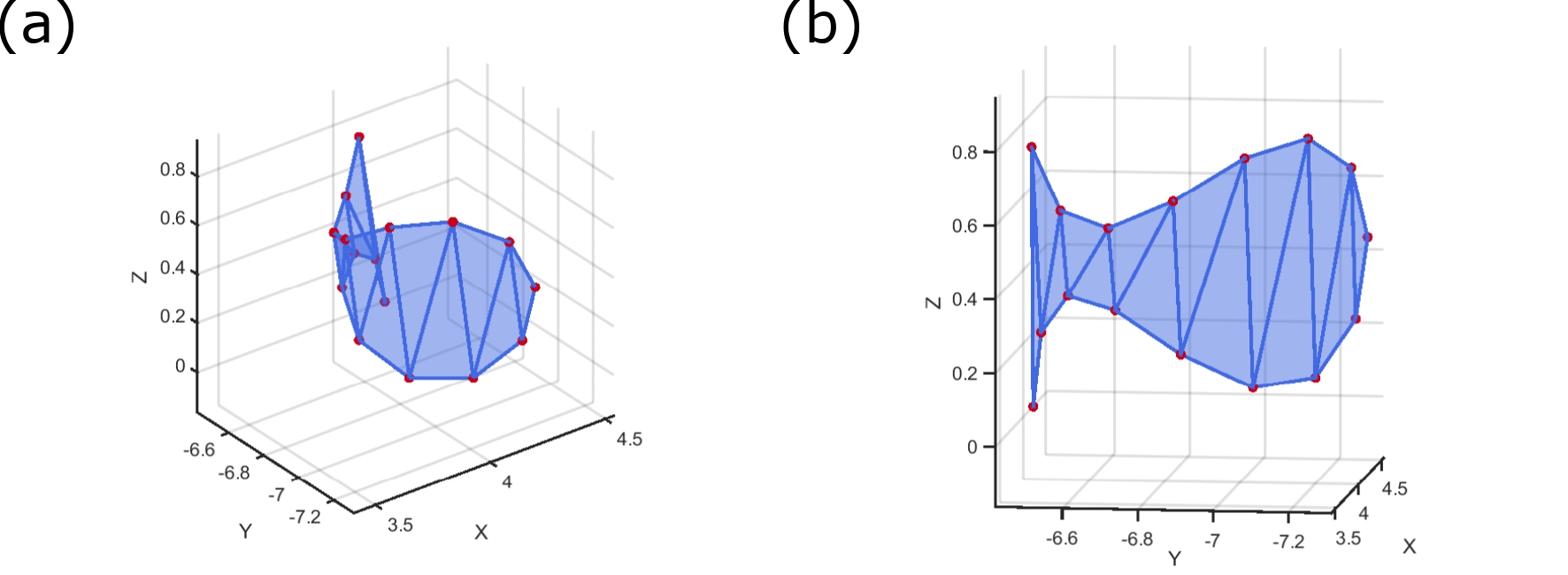}
    \caption{\textbf{Direct triangulation of the given nodes in the unit fabric shape.} Two views are provided in (a) and (b).}
    \label{triangulation}
    \end{figure}

The specific steps of triangulation using 16 nodes are as follows. First, label the 16 nodes in order as $1, 2, 3, \dots, 16$. Then, divide them into two groups of seven triangular patches each, with the triangular patch in the first group has vertex labels: $\{ i, i+1, 17-i \}$ for $i = 1,2,\dots,7$, and those in the second group has vertex labels: $ \{ i, i+1, 17-i \}$, for $i = 9, 10,\dots,15$. Using the known coordinates of the triangle vertices in each group, calculate the area of each triangular patch using the vector dot product, and summing them provides a coarse approximation of the unit area. Fig.~\ref{triangulation} demonstrates the triangulation described above.

\subsection{B\'ezier surface}
As mentioned in the main text, another approach for generating the unit surface is to consider the B\'ezier surface. First, we define four sets of three-dimensional control points $ \{P_{00}, P_{01}, P_{02}, P_{03} \}$, $\{P_{10}, P_{11}, P_{12}, P_{13} \}$, $\{P_{20}, P_{21}, P_{22}, P_{23}\}$, $\{P_{30}, P_{31}, P_{32}, P_{33}\}$, each with four points. Then, a B\'ezier curve from each control point set is generated. Specifically, using the input coordinates of four points, a B\'ezier curve of degree $n = 3$ can be constructed by computing the Bernstein coefficients as basis function coefficients and creating a parameter $u$ ranging from 0 to 1 with the step size $\Delta u$ (we use $\Delta u=0.002$ in our computation). Then, computing the corresponding Bernstein basis function value for each $u$ and concatenating the results into a matrix, the resulting matrix can then represent the desired curve, which is composed of a series of discrete points. 

Here, the B\'ezier curve $P(u)$ is defined by control points $ P_i $ and Bernstein basis functions $ B_i^n(u)$:
\begin{equation}
P(u) = \sum_{i=0}^n B_i^n(u)P_i \quad (u \in [0,1]),
\end{equation}
where the Bernstein basis functions are given by:
\begin{equation}
B_i^n(u) = \binom{n}{i} u^i (1-u)^{n-i}.
\end{equation}
Using the resulting four curves, a surface control mesh could be constructed. Specifically, one can traverse four B\'ezeir curves simultaneously by selecting one point from each curve at a time, generating a new set of four points as control points, from which new B\'ezier curves can be constructed. Iteratively, all points on the original B\'ezier curve will be used once to construct the surface mesh intersecting the original ones. Hence, the resulting B\'ezier surface is composed of a doubly parameterized curve mesh, expressed as:
    \begin{equation}
    S(u,v) = \sum_{i=0}^m \sum_{j=0}^n B_i^m(u)B_j^n(v)P_{ij},
    \end{equation}
where $u,v \in [0,1]$. In practice, we can use the \texttt{bezret} and \texttt{bez3d} functions in MATLAB~\cite{Bezier} to compute the B\'ezier surface based on the above-mentioned procedure.

\begin{figure}[t]
    \centering 
    \includegraphics[width=\linewidth]{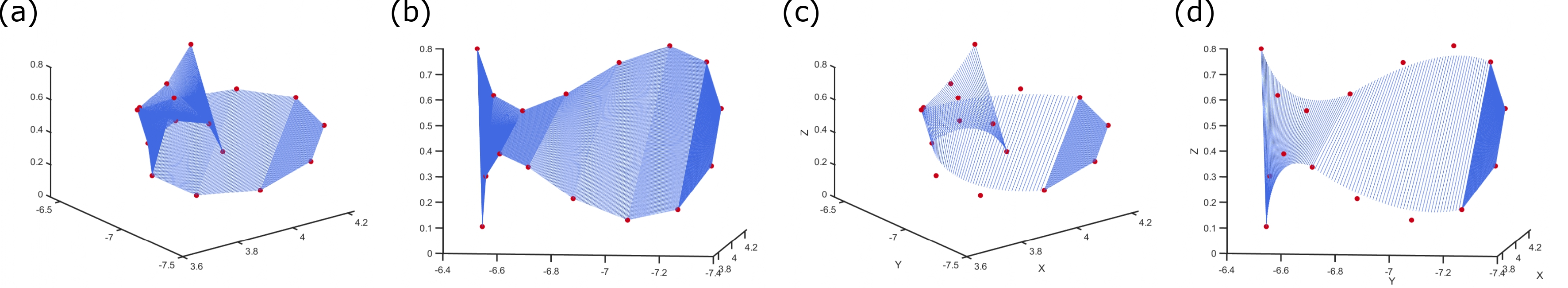}
    \caption{\textbf{Surface construction using the B\'ezier surface.} (a)--(b) B\'ezier surface constructed using seven groups of four points. (c)--(d) B\'ezier surface constructed using two groups of eight points and one group of four points.}
    \label{bezier}
    \end{figure}

Fig.~\ref{bezier}(a)--(b) show the B\'ezier Surface constructed using seven groups of vertices: $\{ i, i + 1, 16 - i, 17 - i \}$, for $i = 1, 2, \dots, 7$. Fig.~\ref{bezier}(c)--(d) show the B\'ezier Surface constructed using three groups of vertices: $\{ 1, 2, 3, 4, 13, 14, 15, 16 \}$, $\{ 4, 5, 6, 7, 10, 11, 12, 13 \}$, and $\{ 7, 8, 9, 10 \}$.

Although the method above is intuitive and computationally simple, it is not precise enough. If too many groups are formed, the B\'ezier surface will be coarse (Fig.~\ref{bezier}(a)--(b)), while if a group contains too many original nodes, the surfaces generated exhibit deviation from the nodes (Fig.~\ref{bezier}(c)--(d)). This motivates us to consider an improved B\'ezier surface construction approach, which will be explained in the next section.

\subsection{Improved B\'ezier Surface}
 We first parametrize the input $(u,v)$ coordinates by linearly mapping them to $[0,1]\times[0,1]$ interval:
    \begin{equation}
    u' = \frac{u - u_{\min}}{u_{\max} - u_{\min}}, \quad
    v' = \frac{v - v_{\min}}{v_{\max} - v_{\min}}.
    \end{equation}
    Next, we construct the basis function matrix by computing 16 bicubic Bernstein basis functions, defined as:
    \begin{equation}
    B_{i,n}(u) = \binom{n}{i} u^i (1-u)^{n-i},\label{bicubicB}
    \end{equation}
where $\binom{n}{i}$ is the binomial coefficient with $n=3$ for cubic B\'ezier curves, $i,j \in [0,3]$ for surface control points, and $u,v \in [0,1]$ parameterize the unit domain.

The bicubic form combines two basis functions:
    \begin{equation}
    B_{i,j}(u,v) = B_{i,3}(u)  B_{j,3}(v).
    \end{equation}
Subsequently, we can determine the new 16 control points from the original ones by constructing an overdetermined system of equations and solving for the least-squares solution using the pseudo-inverse matrix. Specifically, with the original set of nodal points, which has been parametrized as $(u_i,v_i)$ with corresponding surface points $\bm{P}_i$, we construct the overdetermined system:
\begin{equation} 
\bm{B} \bm{K} = \bm{P},
\end{equation}
where $\bm{B}$ is the $m \times 16$ basis function matrix ($m \geq 16$), $\bm{K}$ is the $16 \times 3$ control point matrix, and $\bm{P}$ is the $m \times 3$ point matrix. The least-squares solution for the above linear system is obtained using the Moore-Penrose pseudoinverse:
\begin{equation} 
\bm{K} = \bm{B}^+ \bm{P}, 
\end{equation}
where the pseudoinverse is defined as:
\begin{equation} 
\bm{B}^+ = (\bm{B}^T \bm{B})^{-1} \bm{B}^T.
\end{equation}
Note that the above procedure is performed independently in three dimensions:
\begin{equation} 
\bm{K}_x = \bm{B}^+ \bm{P}_x \ , \quad \bm{K}_y = \bm{B}^+ \bm{P}_y \ , \quad \bm{K}_z = \bm{B}^+ \bm{P}_z.
\end{equation}
In practice, we use the MATLAB implementation as described in~\cite{choi2019programming} to construct and solve the above-mentioned system.

    \begin{figure}[t]
    \centering 
    \includegraphics[width=0.8\linewidth]{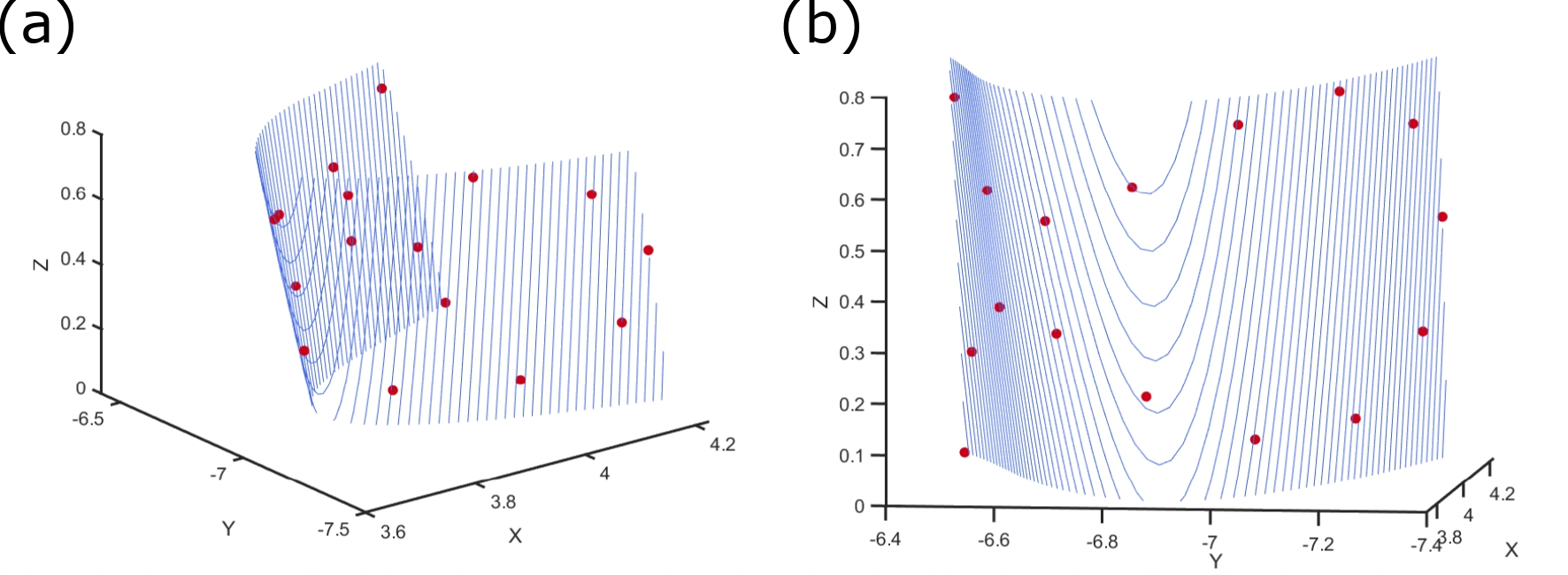}
    \caption{\textbf{Surface construction using improved B\'ezier surface.} Two views are provided in (a) and (b).}
    \label{bezier new}
    \end{figure}
    
Fig.~\ref{bezier new} shows that the issue of excessive deviation between the surface and control points has been resolved. However, it also demonstrates the problem that the point cloud generated by the above code may extend beyond the unit surface area.
   
To compute the Delaunay triangulation, the points in the point cloud are first projected onto a certain plane (in our example, the $x$-$y$ plane). Following this, the Delaunay triangulation will give a triangulation of the projected points, whose boundary is formed by the points on the convex hull of the projected points. Meanwhile, the triangulation will make the smallest interior angle of all triangles as large as possible to avoid elongated triangles. Then, compute the area of the triangular patches with original node positions, and summing up will give the surface area of the convex hull containing the point cloud. 

\begin{figure}[t]
    \centering 
    \includegraphics[width=\linewidth]{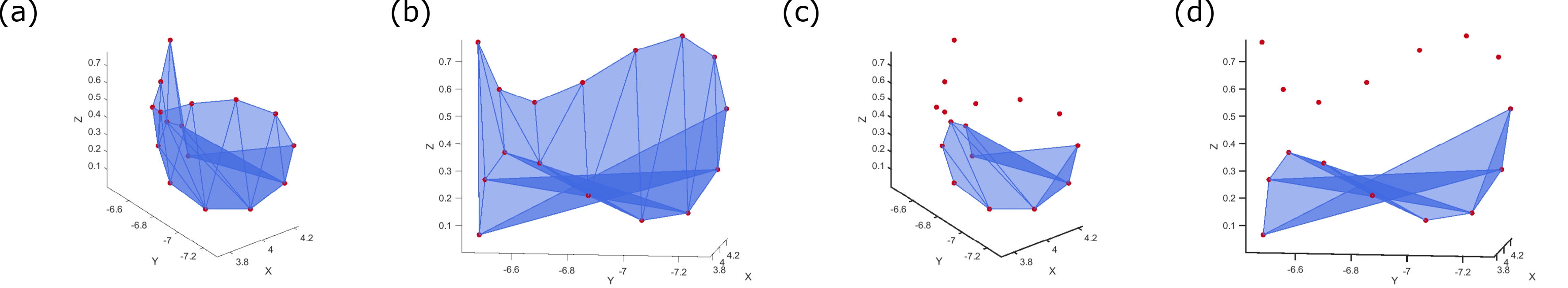}
    \caption{\textbf{Surface obtained using Delaunay triangulation.} (a)--(b) Surface obtained using Delaunay triangulation of projected 16 nodes. (c)--(d) Surface obtained using Delaunay triangulation of projected boundary points of the unwanted bottom surface.}
    \label{delaunay}
    \end{figure}

Fig.~\ref{delaunay} illustrates a simple example of a Delaunay triangulation, which fails to produce the correct surface we desire. Using the 16 nodal points as an example, the Delaunay triangulation method will return a surface that includes the side faces we do not need. However, if we attempt to manually remove this side face, Delaunay triangulation will then return a surface containing triangular faces that we do not intend to remove.

\subsection{Ruled surface}

To construct a ruled surface from the reconstructed boundary curve, we select $n_b$ points on the boundary curve.

First, we label the $n_b$ points on the obtained boundary curve in order as $1, 2, \dots, i, \dots, n_b$, and denote their coordinates by $\Vec{p}(i)$. We then parametrize the ruled surface $S$ by $(m, n)$ for $m\in \{1,2,\dots, \frac{n_b}{2}\}$, $n \in \{1,2,\dots, n_l\}$, and compute $S(m, n)$ by:
\begin{equation}
   S(m, n)= \Vec{\alpha} (m)+ \frac{n-1}{n_l-1}\Vec{\beta}(m),
   \end{equation}
where $\Vec{\alpha}(m)= \Vec{p}(m)$ and $\Vec{\beta}(m)=\Vec{p}(n_b-m+1)-\Vec{p}(m)$. In this way, we obtain a mesh of $n_b \times n_l$ sampling points on the ruled surface with the boundary curve being the given B-spline curve.

\begin{figure}[t]
    \centering 
    \includegraphics[width=\linewidth]{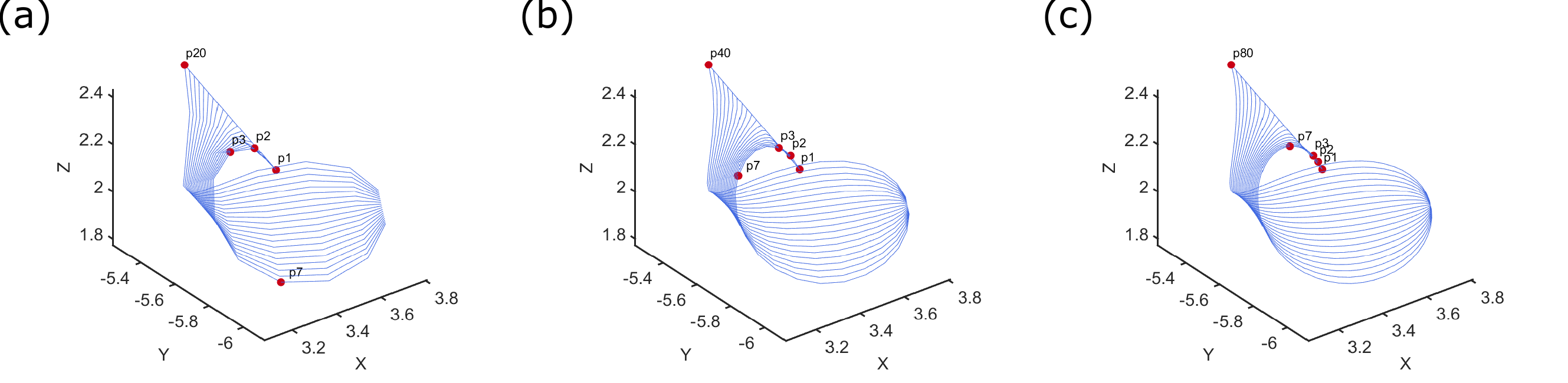}
    \caption{\textbf{Surface reconstruction obtained by combining the B-spline boundary curve and the ruled surface.} (a) Ruled surface using 20 boundary points. (b) Ruled surface using 40 boundary points. (c) Ruled surface using 80 boundary points. }
    \label{ruledsurface}
    \end{figure}

 With larger $n_b$ and $n_l$, a smoother surface could be plotted. For example, the surface in Fig.~\ref{ruledsurface}(a) uses 20 boundary points to reconstruct the boundary curve, and each generator thereafter uses 20 control points. Whereas the surface in Fig.~\ref{ruledsurface}(b) increased the number of boundary points to 40, and that in Fig.~\ref{ruledsurface}(c) increased to 80, maintaining the number of control points on each generator unchanged. By comparing the three surfaces, it is obvious that the surface with more control points is smoother.

\subsection{Unit area quantification}

Using the above-mentioned surface reconstruction methods, we can easily quantify the area of each unit cell.

\begin{figure}[t]
    \centering 
    \includegraphics[width=0.8\linewidth]{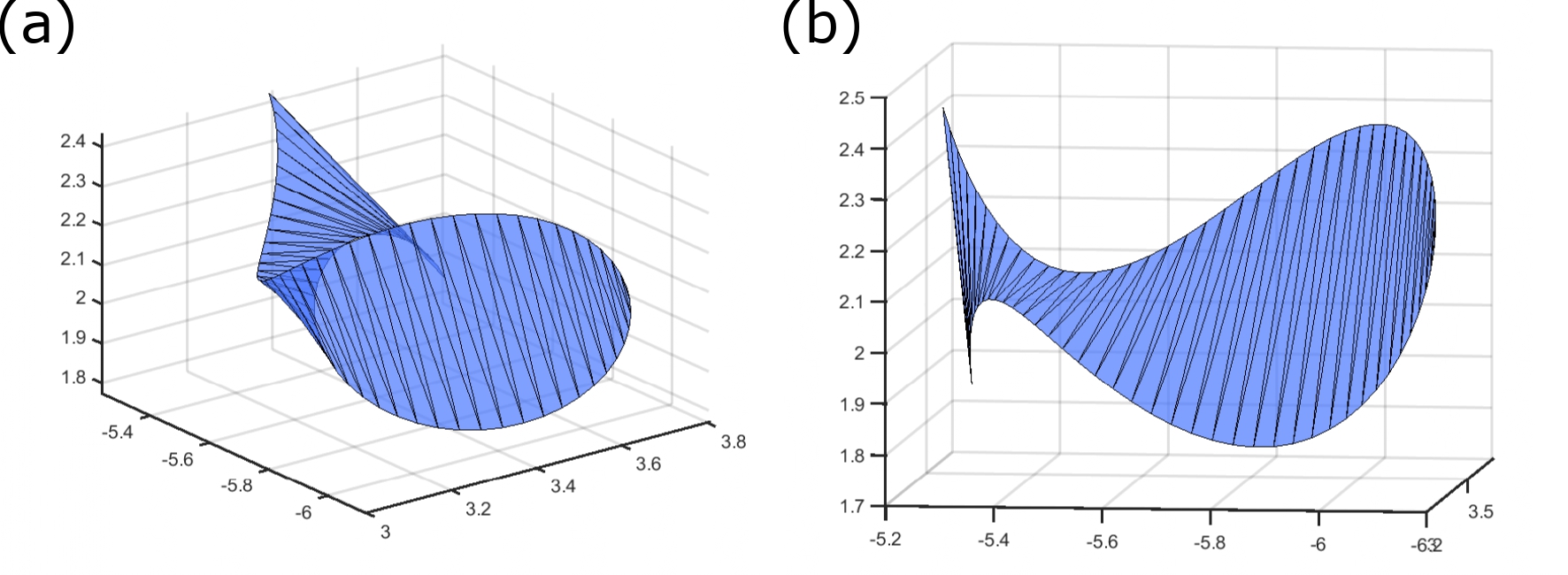}
    \caption{\textbf{Triangulation constructed using the B-spline boundary curve for unit area quantification.} Two views are provided in (a) and (b).}
    \label{ruledarea}
    \end{figure}

For the triangulation approach, we can compute the area by directly summing up the area of each triangle element. As for computing the surface area using the B-spline boundary curve, we can start from the $n_b$ sampling points on the boundary curve and perform a triangulation using each pair of adjacent points on the same side of the boundary and the point on the opposite side. For example, connecting points of index $i$, $i+1$, and the symmetrical point of index $n_b+1-i$ can form a triangle located on the upper part of the surface; connecting points of index $i$, $i-1$, and the symmetrical point of index $n_b+2-i$ can form a triangle located on the lower part of the surface. Fig.~\ref{ruledarea} demonstrates the triangulation result using the above methods. After forming the triangulation, we can compute the unit area by adding up all the individual triangle areas.

\subsection{Surface curvature quantification}

 Through the ruled surface reconstruction, we can obtain $n_b \times n_l$ discrete points on the surface. The $X, Y, Z$ coordinates of these points can be stored in three two-dimensional arrays $X$, $Y$, $Z$ of size $n_b$ by $n_l$ (for example, $X(i,j)$ corresponds to the x-coordinate of the $(i,j)$-th point). Values such as $X_u$, $X_v$, $X_{uu}$, $X_{uv}$ can be calculated using the gradient function in MATLAB. With partial derivatives of $X$, $Y$ and $Z$ along the $u$ and $v$ directions, coefficients of the first and second fundamental form $E$, $F$, $G$, $e$, $f$, $g$ of the surface at discrete points can be calculated accordingly, and $K$ (Gaussian curvature) and $H$ (Mean curvature) follow from the formulas:
\begin{equation}
K = \frac{eg - f^2}{EG - F^2},
\end{equation}   
and
\begin{equation}
H = \frac{eG - 2fF + gE}{2(EG - F^2)}.
\end{equation}

\subsection{Comparison between different surface quantification methods}

To assess the suitability of the above-mentioned surface reconstruction methods for quantifying the surface-based quantities of the fabric structures, we performed a detailed comparison between them in terms of the area and surface curvature obtained from them.

First, we compared the resulting unit surface area in Table~\ref{comparison_surface_area}. Among all the methods, the hybrid technique combining B-spline and ruled surface is the most satisfactory in balancing the accuracy and efficiency. 

 \begin{table}[t]
\centering
\begin{tabular}{|C{30mm}|C{20mm}|C{25mm}|C{25mm}|C{40mm}|}
\hline
\textbf{Method} & \textbf{Number of Points} & \textbf{Unit Area} & \textbf{Percentage Difference} & \textbf{Remark} \\
\hline
Monte-Carlo method & 1300000 & 0.5876  & - &Large computational cost \\ \hline
Ruled surface & 1000 & 0.5718 & 2.6889\% & More efficient with decent accuracy \\ \hline
Direct triangulation & 16 & 0.5558 &5.4118\% & Unsatisfying accuracy \\
\hline  
\end{tabular}
\caption{Comparison between different surface reconstruction methods for surface area quantification.} 
\label{comparison_surface_area}
\end{table}

Next, we compare the surface curvature calculated by different surface mesh construction methods (Table~\ref{comparison_surface_curvature}). When using the B\'ezier surface with seven groups of four points, the range of each surface will be small. Therefore, the resulting surfaces will be more twisted. If two sets of eight points and a set of four points are used to construct the B\'ezier surface, the resulting surface will demonstrate severe deviation from the original nodes. Since a B\'ezier curve generally does not pass through the intermediate points, but instead forms a smoother curve among them, the generated surface is less curved than in reality. Hence, the curvature is smaller than reality. For the case when new control points of the B\'ezier surface are chosen, the resulting surface will not fit well with the boundary curve, thereby including redundant area for curvature calculation. Observing the surface constructed by the improved B\'ezier surface algorithm,  the most curved part (around $y \in [-6.6,-7.1]$) has the most redundant area, which explains the reason for the larger curvature result. In comparison, when the B-spline curve and ruled surface are used to form a surface mesh, the reconstructed surface fits well with the original nodes, with no obvious sharp turnings or twisting, and no redundant area is included. 

\begin{table}[t]
\centering
\begin{tabular}{|C{25mm}|C{20mm}|C{25mm}|C{25mm}|C{40mm}|}
\hline
\textbf{Method} & \textbf{Number of Points} & \textbf{Gaussian Curvature} & \textbf{Mean Curvature} & \textbf{Remark} \\
\hline
B\'ezier surface (7 partitions) & 18207 & 12.1347 & 6.1706 & Sharp turnings (large curvatures) occur at edges \\ \hline
B\'ezier surface (3 partitions) & 7803 & 0.7784 & 0.4216 & large deviation from original nodes \\ \hline
B\'ezier surface (new control points) & 7803 & 6.2852 & 1.1821 & Do not fit the boundary curve (redundant area) \\ \hline
Ruled surface & 400 & 1.7884 & 0.6571 & Use fewer points with an accurate range and a smooth surface \\
\hline  
\end{tabular}
\caption{Comparison between different surface reconstruction methods for surface curvature quantification.}
\label{comparison_surface_curvature}
\end{table}

\section{Volume quantification} 
In this section, we present three approaches we considered for constructing and quantifying the 3D structure of the unit fabric.

\subsection{Axis-aligned bounding box}
First, we considered using axis-aligned bounding boxes for the volume quantification. In this approach, we directly consider approximating the 3D shape using a bounding box aligned with the $x$, $y$, and $z$ axes. We can then obtain the length, width, and height of the bounding box:
    \begin{equation}
    \Delta x = x_{\max}-x_{\min} \quad ,\quad 
    \Delta y = y_{\max}-y_{\min} \quad ,\quad
    \Delta z = z_{\max}-z_{\min},
    \end{equation}
from which we can easily obtain various 3D measurements:
    \begin{equation} 
    \text{Volume} = \Delta x \cdot \Delta y \cdot \Delta z, 
    \end{equation}
    \begin{equation} 
    \text{Aspect ratio} = \frac{\Delta y}{ \Delta x},
    \end{equation}
     \begin{equation} 
    \text{Centroid} = \left[ \frac{x_{\min} + x_{\max}}{2}, \frac{y_{\min} + y_{\max}}{2}, \frac{z_{\min} + z_{\max}}{2} \right].
    \end{equation}

\subsection{Minimal bounding box}
Another approach for the volume quantification is to consider a minimal bounding box, which computes a bounding box with the minimum volume for a set of points without any assumption on the alignment with the coordinate axes. In particular, one can first compute the convex hull of the point cloud to rule out those points inside the convex hull, so that the problem size can be reduced. Following this, a local coordinate system can be established on each convex hull face using the Schmidt orthogonalization method. Subsequently, the point cloud is rotated into each local coordinate system, the corresponding bounding box dimensions are calculated, and then compared with the current optimal solution for updating. After comparing all the local coordinates, the best solution can be obtained. In practice, we obtain the minimum bounding box based on the above procedure using the \texttt{minboundbox} function in MATLAB~\cite{minboundbox}. After obtaining the minimum bounding box, we can further calculate the bounding box volume $v$ and record the coordinates of its 8 corner points. Using the centroid of the minimal bounding box (i.e., average of the coordinates of the eight corner points), the position of the unit fabric is determined. Finally, we calculate the aspect ratio as the ratio of the maximum distance of points in the point cloud along the $x$ and $y$ directions. 

\subsection{Convex hull}

We further consider computing the convex hull for volume quantification. Note that the convex hull of a given set of points is the smallest convex polyhedron that contains all the points. Finding the convex hull typically involves the following steps: First, one has to sort the set of points by polar angle (i.e., the angle with a reference point) to form an ordered sequence. Then, one can select the extreme points that constitute the convex hull boundary among the ordered points. Finally, the point set is divided into multiple subsets, and each subset is processed to approximate the convex hull boundary. In practice, we can utilize the built-in \texttt{convhulln} function in MATLAB to obtain the convex hull together with its volume. The position of each unit fabric is defined as the average of the extreme values of the coordinates of points in the point set:  $\text{Centroid} = \left[ \frac{x_{\min} + x_{\max}}{2}, \frac{y_{\min} + y_{\max}}{2}, \frac{z_{\min} + z_{\max}}{2} \right]$. The aspect ratio is defined the same as above: $ \text{Aspect ratio} = \frac{\Delta y}{ \Delta x}$.

\subsection{Comparison between different volume quantification methods}

\begin{figure}[t]
    \centering
     \includegraphics[width=1\linewidth]{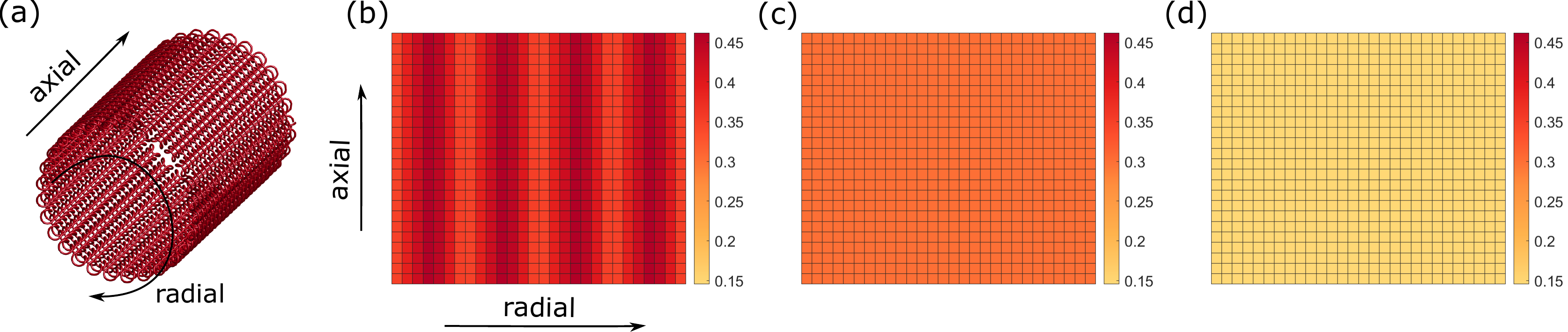}
    \caption{\textbf{Comparison between different volume quantification methods.} (a) A cylindrical sample used for the comparison. (b)--(d) We calculated the unit volume using three methods: (b) axis-aligned bounding box, (c) minimal bounding box, (d) convex hull.}
     \label{SI.Compare_3_methods_for_v}
\end{figure}

We tested three volume quantification methods on a cylindrical sample (Fig.~\ref{SI.Compare_3_methods_for_v}(a)) to evaluate their capability of handling 3D structures.

We calculated the unit volume at a static state, and the resulting data demonstrated obviously different patterns (Fig.~\ref{SI.Compare_3_methods_for_v}(b)--(d)). The data obtained using the axis-aligned bounding box illustrates a periodic pattern along the radial direction, which could be explained by the curling structure of the cylinder, which leads to the continuous rotation of the unit fabric. When a unit fabric is oriented parallel to a coordinate plane (e.g., the x-y plane), the three dimensions of the bounding box closely conform to the actual extent of the stitch, minimizing redundant space and occupying volume that closely approximates the true material occupancy. In contrast, when the unit stitch is aligned along a diagonal or oblique plane of the axis-aligned bounding box, the axis-aligned bounding box must encompass its maximum projections across all coordinate axes, resulting in significant void regions. This leads to ‌overestimation of volume‌, which can compromise modeling accuracy and adversely affect subsequent mechanical analyses. When using the unit volume computed by the minimal bounding box (Fig.~\ref{SI.Compare_3_methods_for_v}(c)) and convex hull (Fig.~\ref{SI.Compare_3_methods_for_v}(d)), results illustrated a consistent pattern throughout the sampling fabric, confirming the previous discussion that these two methods are well-suited for spatial rotation. Moreover, the maximal unit volume computed by the axis-aligned bounding box, the minimal bounding box, and the convex hull are 0.35, 0.3, and 0.15, respectively, which is also consistent with previous assertions that the convex hull is the most effective in adapting to complex spatial structures and reducing redundant space.

Table~\ref{comparison3vmethods} summarizes the characteristics of different volume calculation methods.

\begin{table}[t]
\centering
\begin{tabular}{c|c|C{25mm}|c}
\hline
\textbf{Method} & \textbf{Results} & \textbf{Orientation Invariance} & \textbf{Ratio to Minimum Value}\\
\hline
Axis-aligned bounding box & 0.3486-0.4613 & No & 239.92\%-317.48\%\\
Minimal bounding box & 0.2999 & Yes & 206.40\%\\
Convex hull & 0.1453 & Yes & 100\%\\
\hline  
\end{tabular}
\caption{Comparison of distances between surfaces generated by different surface construction methods and nodes.}
\label{comparison3vmethods}
\end{table}

\section{Additional results}

\subsection{Anisotropic mechanical responses}

As discussed in the main text, after developing our geometric quantification framework, we can analyze the anisotropic mechanical responses for different fabric structures. In the main text, we presented the results for the jersey pattern. 

\begin{figure}[t]
    \centering
     \includegraphics[width=1\linewidth]{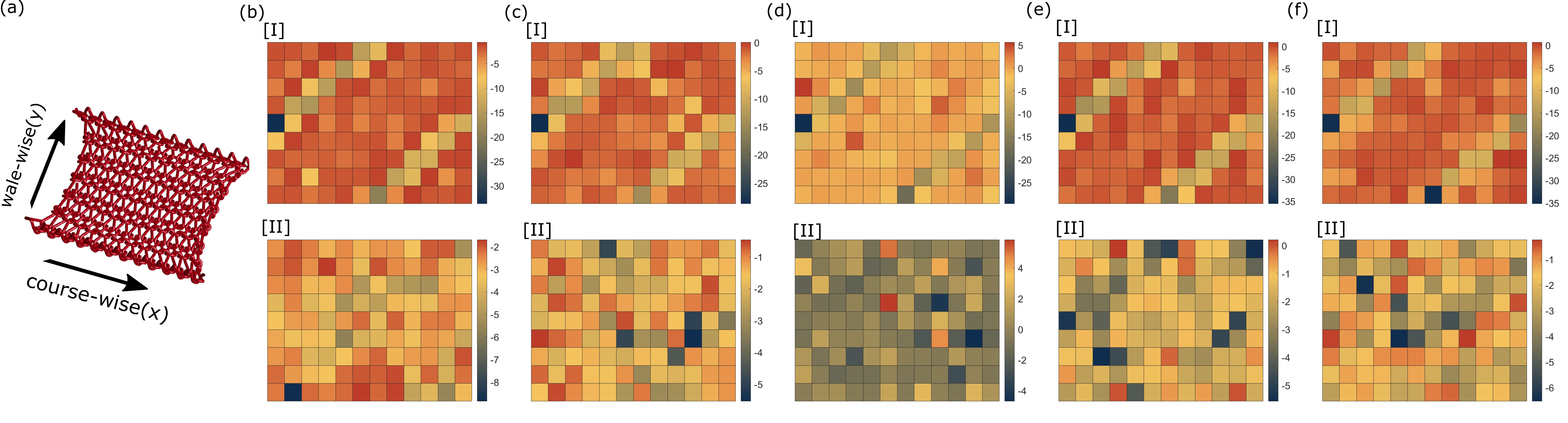}
    \caption{\textbf{Quantification of the anisotropic mechanical responses in the garter pattern.} (a) The garter pattern. (b)--(f) {Spatial distribution of temporal changes in five representative quantities under tensile strain from 0\% to 120\%, where the color of each cell denotes the total change in the geometric quantity for the corresponding stitch:} (b)~Curve curvature, (c)~Curve torsion, (d)~Gaussian curvature, (e) Area, (f)~Volume. Loading directions: [I]~{Course-wise} tension. [II]~{Wale-wise} Tension.}
     \label{fig2.SI.garter}
\end{figure}

\begin{figure}[t!]
    \centering
    \includegraphics[width=1\linewidth]{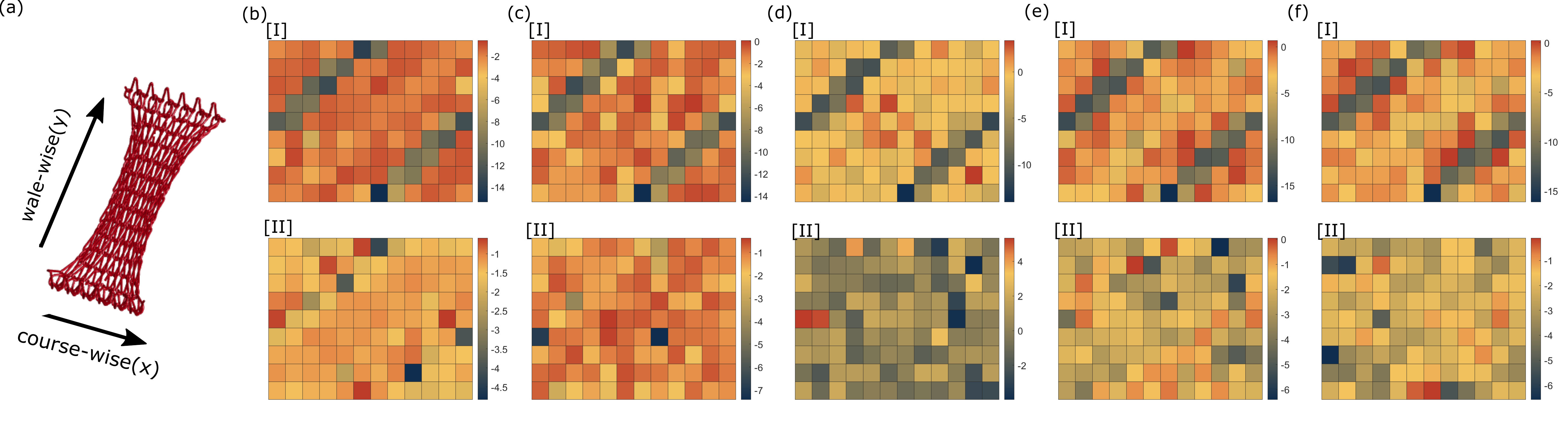}
    \caption{\textbf{Quantification of the anisotropic mechanical responses in the rib pattern.} (a) The rib pattern. (b)--(f) { Spatial distribution of temporal changes in five representative quantities under tensile strain from 0\% to 120\%, where the color of each cell denotes the total change in the geometric quantity for the corresponding stitch:} (b)~Curve curvature, (c)~Curve torsion, (d)~Gaussian curvature, (e) Area, (f)~Volume. Loading directions: [I]~{Course-wise} Tension. [II]~{Wale-wise} Tension.}
     \label{fig2.SI.rib}
\end{figure}

\begin{figure}[t!]
    \centering
     \includegraphics[width=1\linewidth]{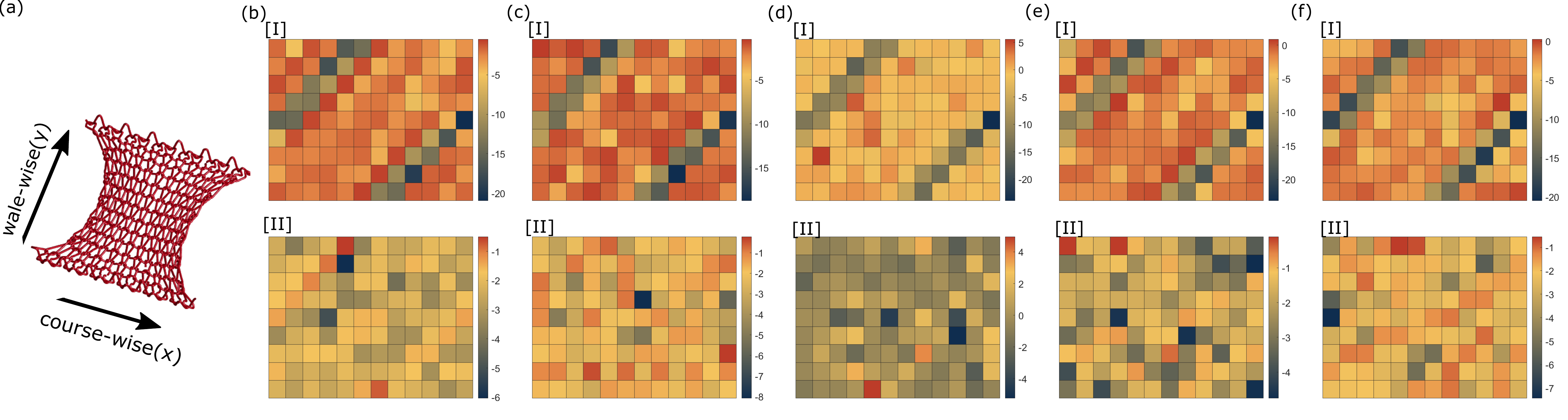}
    \caption{\textbf{Quantification of the anisotropic mechanical responses in the seed pattern.} (a) The seed pattern. (b)--(f) { Spatial distribution of temporal changes in five representative quantities under tensile strain from 0\% to 120\%, where the color of each cell denotes the total change in the geometric quantity for the corresponding stitch:} (b)~Curve curvature, (c)~Curve torsion, (d)~Gaussian curvature, (e) Area, (f)~Volume. Loading directions: [I]~{Course-wise} tension. [II]~{Wale-wise} tension.}
     \label{fig2.SI.seed}
\end{figure}

In Fig.~\ref{fig2.SI.garter}--\ref{fig2.SI.seed}, we further show the quantification of the anisotropic mechanical responses in the garter, rib, and seed patterns.

\subsection{Heterogeneous mechanical responses}

\begin{figure}[t!]
     \centering
     \includegraphics[width=1\linewidth]{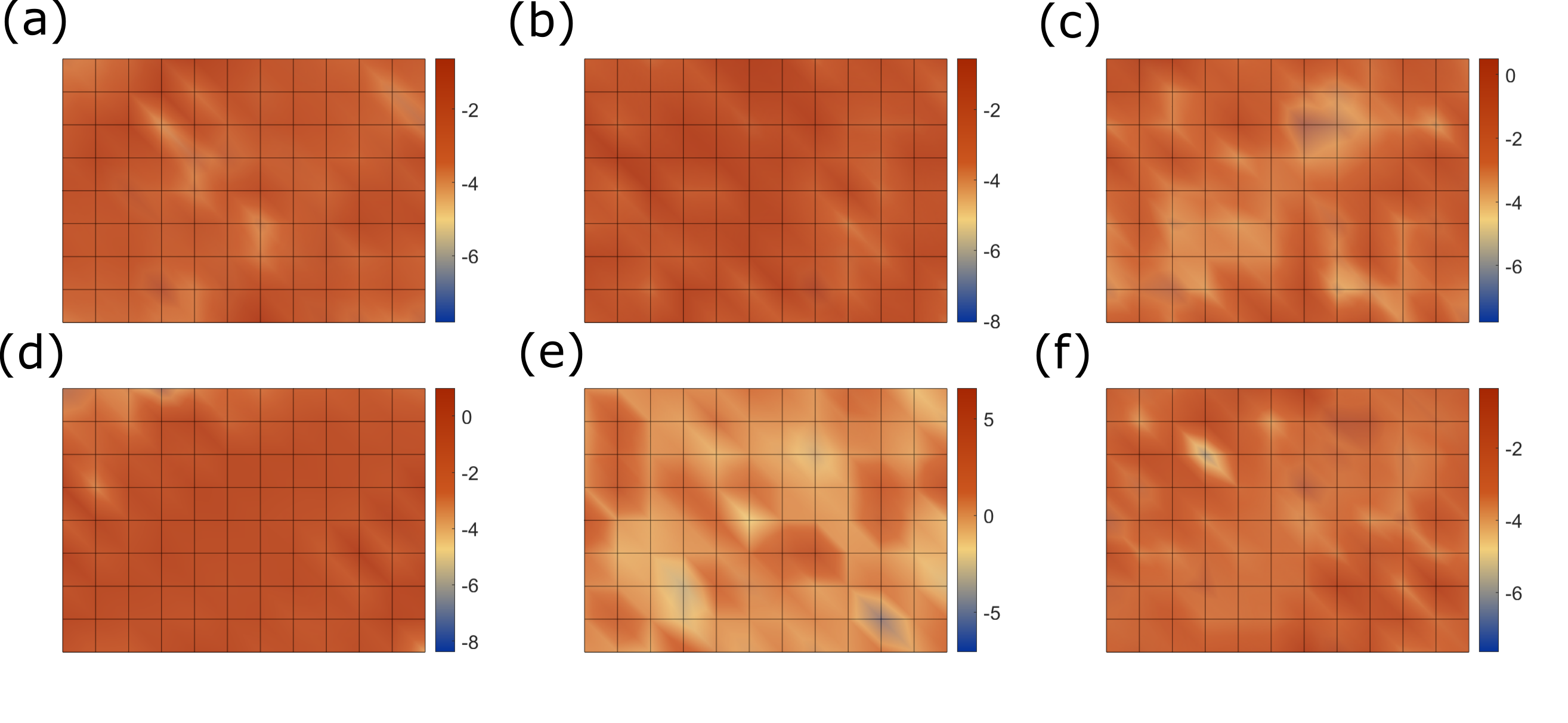}
     \caption{\textbf{Heterogeneous mechanical responses and hot spots of representative quantities in jersey fabrics under {wale-wise} tension.} Time-aggregated maps of changes in six key geometric quantities, generated by overlaying normalized heat maps acquired during successive tensile loading at 0\%, 20\%, 40\%, 60\%, 80\%, 100\%, and 120\% strain with applied spatial smoothing: (a)~Curve curvature. (b)~Curve torsion. (c)~Area. (d)~Gaussian curvature. (e)~Mean curvature. (f)~Aspect ratio. (g)~Volume.}
     \label{fig3.SI.jersey.smooth}
 \end{figure}

\begin{figure}[t!]
     \centering
     \includegraphics[width=1\linewidth]{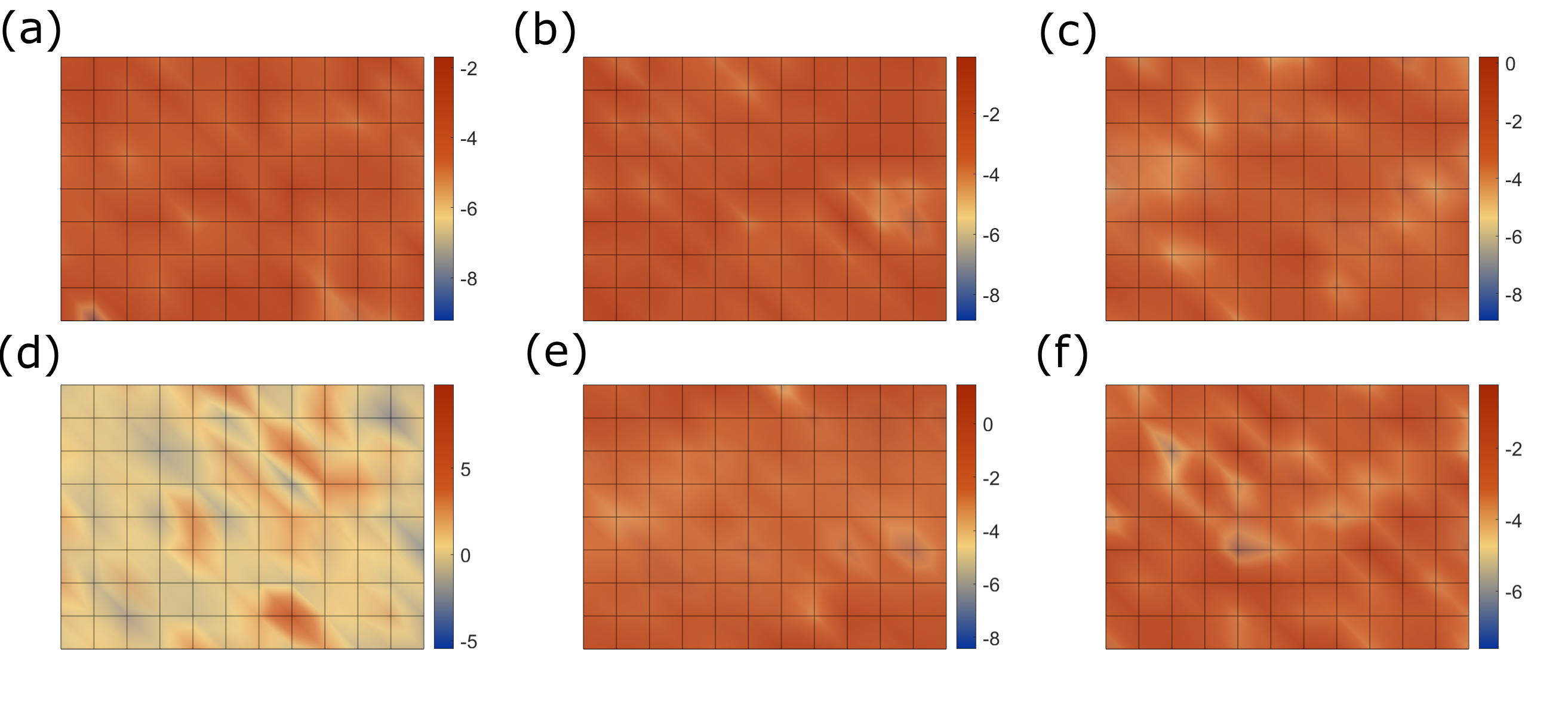}
     \caption{\textbf{Heterogeneous mechanical responses and hot spots of representative quantities in garter fabrics under {wale-wise} tension.} Time-aggregated maps of changes in six key geometric quantities, generated by overlaying normalized heat maps acquired during successive tensile loading at 0\%, 20\%, 40\%, 60\%, 80\%, 100\%, and 120\% strain with applied spatial smoothing: (a)~Curve curvature. (b)~Curve torsion. (c)~Area. (d)~Gaussian curvature. (e)~Mean curvature. (f)~Aspect ratio. (g)~Volume.}
     \label{fig3.SI.garter.smooth}
 \end{figure}

\begin{figure}[t!]
     \centering
     \includegraphics[width=0.95\linewidth]{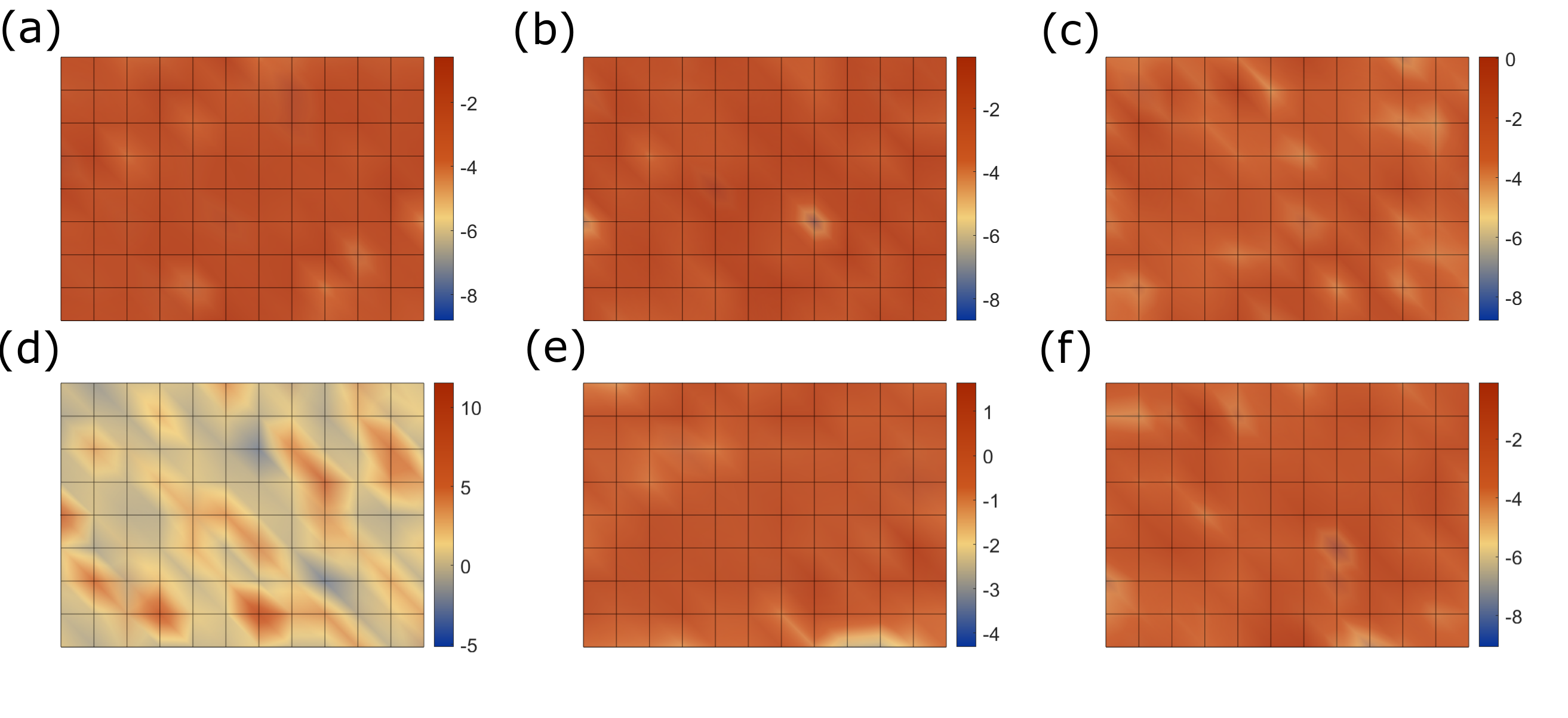}
     \caption{\textbf{Heterogeneous mechanical responses and hot spots of representative quantities in rib fabrics under {wale-wise} tension.} Time-aggregated maps of changes in six key geometric quantities, generated by overlaying normalized heat maps acquired during successive tensile loading at 0\%, 20\%, 40\%, 60\%, 80\%, 100\%, and 120\% strain with applied spatial smoothing: (a)~Curve curvature. (b)~Curve torsion. (c)~Area. (d)~Gaussian curvature. (e)~Mean curvature. (f)~Aspect ratio. (g)~Volume.}
     \label{fig3.SI.rib.smooth}
 \end{figure}

\begin{figure}[t!]
     \centering
     \includegraphics[width=0.95\linewidth]{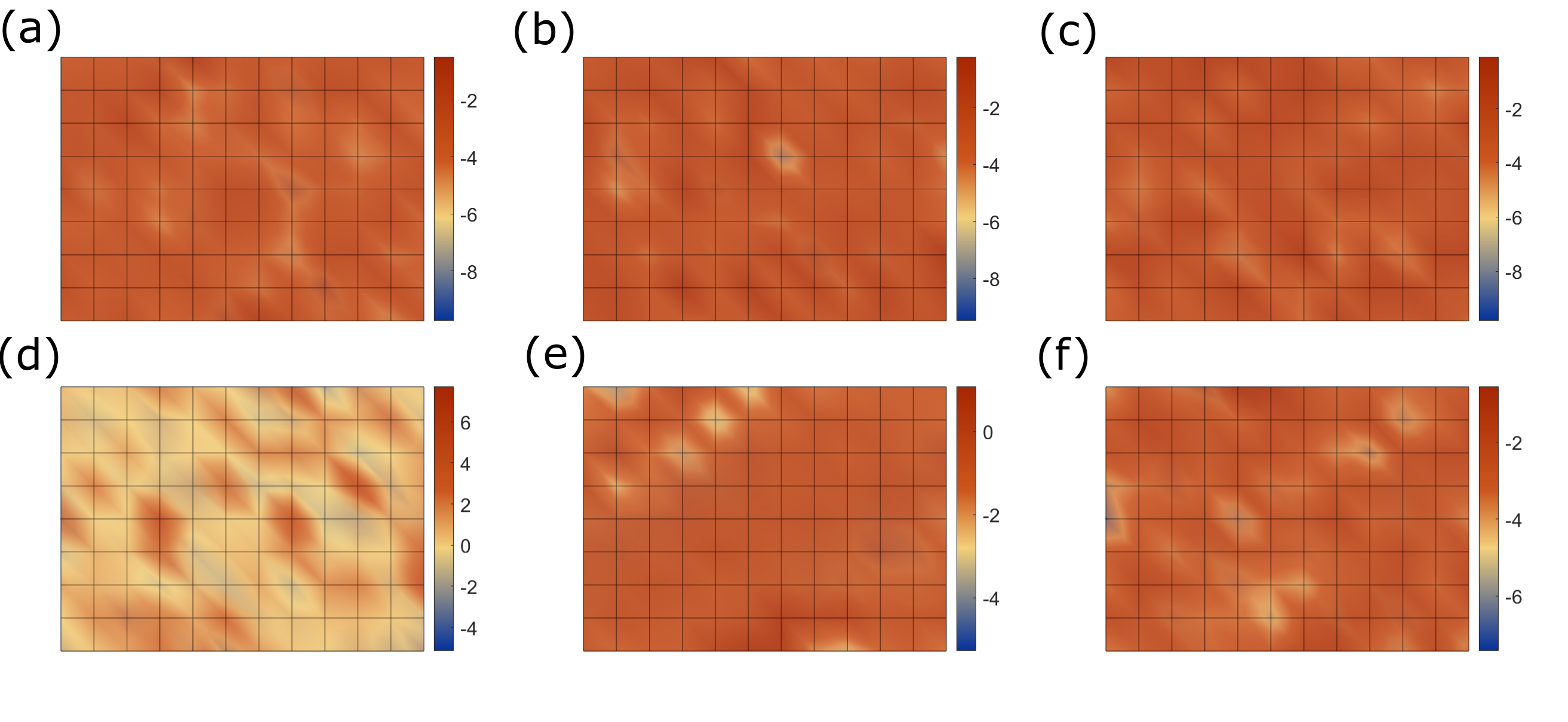}
     \caption{\textbf{Heterogeneous mechanical responses and hot spots of representative quantities in seed fabrics under {wale-wise} tension.} Time-aggregated maps of changes in six key geometric quantities, generated by overlaying normalized heat maps acquired during successive tensile loading at 0\%, 20\%, 40\%, 60\%, 80\%, 100\%, and 120\% strain with applied spatial smoothing: (a)~Curve curvature. (b)~Curve torsion. (c)~Area. (d)~Gaussian curvature. (e)~Mean curvature. (f)~Aspect ratio. (g)~Volume.}
     \label{fig3.SI.seed.smooth}
 \end{figure}

In Fig.~\ref{fig3.SI.jersey.smooth}--\ref{fig3.SI.seed.smooth}, we show hot spots of local geometric quantities in four patterns. This is formed by overlaying heatmaps of temporal changes in the data with a certain level of transparency and by smoothing out boundaries between units to simulate realistic transitions between regions. We can therefore observe the uniformity of certain changes in quantity and detect regions undergoing the most intensive change.

\subsection{Mechanical responses from mixed-pattern fabrics}

\begin{figure}[t!]
    \centering
    \includegraphics[width=0.8\linewidth]{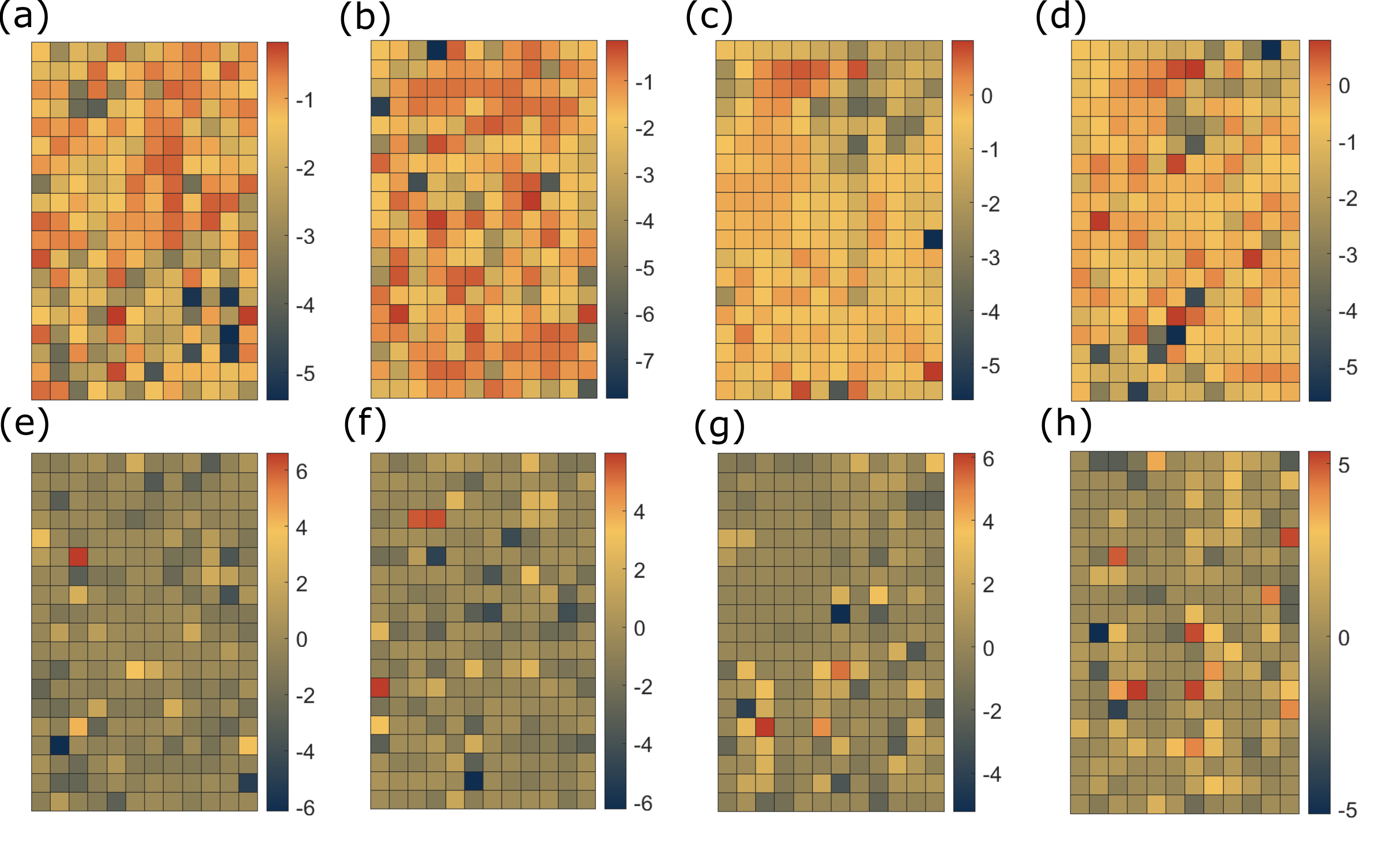}
    \caption{\textbf{Quantifying the mechanical responses from mixed-pattern fabrics.}  Spatial distribution of geometric quantity changes in two mixed-pattern fabrics under {wale-wise} uniaxial tensile strain from 0\% to 180\%: (a)~Curve torsion change for pattern A. (b)~Curve torsion change for pattern B. (c)~Area change for pattern A. (d)~Area change for pattern B. (e)~Gaussian curvature change for pattern A. (f)~Gaussian curvature change for pattern B. (g)~Mean curvature change for pattern A. (h)~Mean curvature change for pattern B.}
     \label{fig4.SI}
\end{figure}

As for the mixed-pattern fabrics, in the main text we presented the analysis of their mechanical responses in terms of several geometric quantities. In particular, we considered two mixed-pattern fabrics:
\begin{itemize}
    \item Pattern A (main text Fig.~4(a)): A mixed pattern with the top half being jersey and the bottom half being garter. 
    \item Pattern B (main text Fig.~4(b)): A mixed pattern with the top and bottom quarter being jersey and the middle half being garter.
\end{itemize}
In Fig.~\ref{fig4.SI}, we further present other geometric quantities of both Pattern A and Pattern B, including the curve torsion variation, area variation, Gaussian curvature variation, and mean curvature variation.

\subsection{Temporal change of the spatial variation of 2.5D fabrics: Onset and evolution of hot spots}

\begin{figure}[t]
     \centering
     \includegraphics[width=1\linewidth]{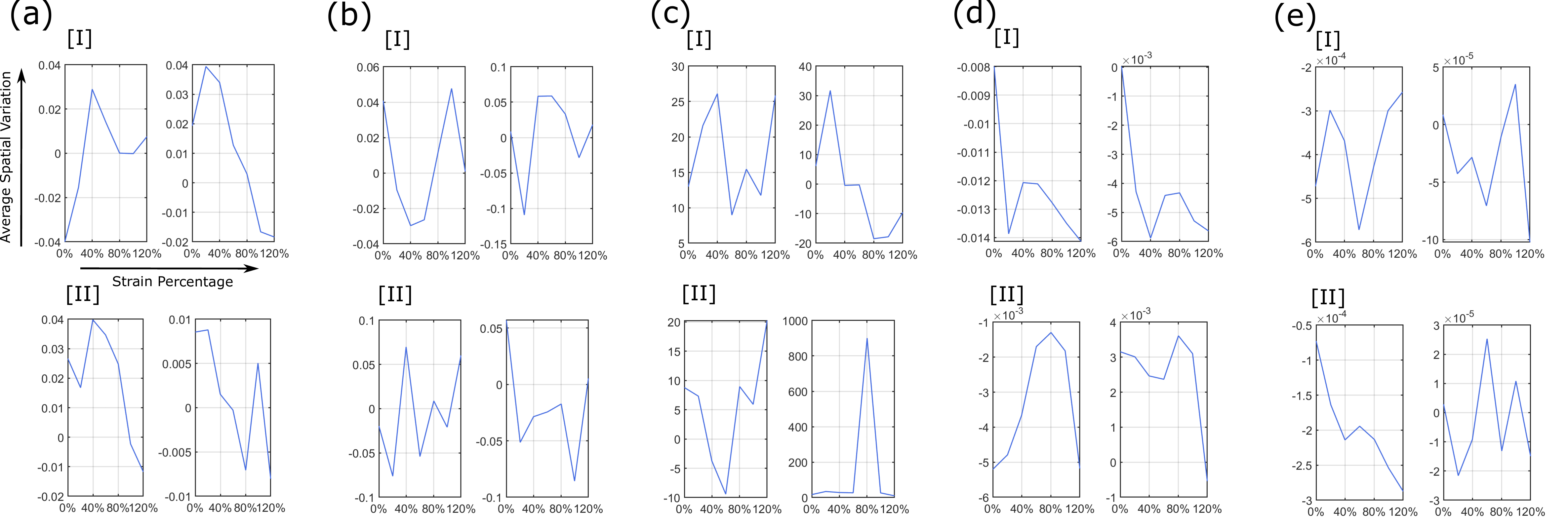}
     \caption{\textbf{Temporal changes in spatial variations of four representative quantities for garter fabrics, recorded across a tensile strain range of 0\% to 120\%.} (a)~Curve curvature. (b)~Curve torsion. (c)~Gaussian curvature. (d)~Area. (e)~Volume. Loading directions: [I]~{Course-wise}. [II]~{Wale-wise}. In each of (a)--(e), the left plot corresponds to {course-wise} variation and the right plot corresponds to {wale-wise} variation.}
     \label{fig5.SI.garter}
 \end{figure}

\begin{figure}[t]
     \centering
     \includegraphics[width=1\linewidth]{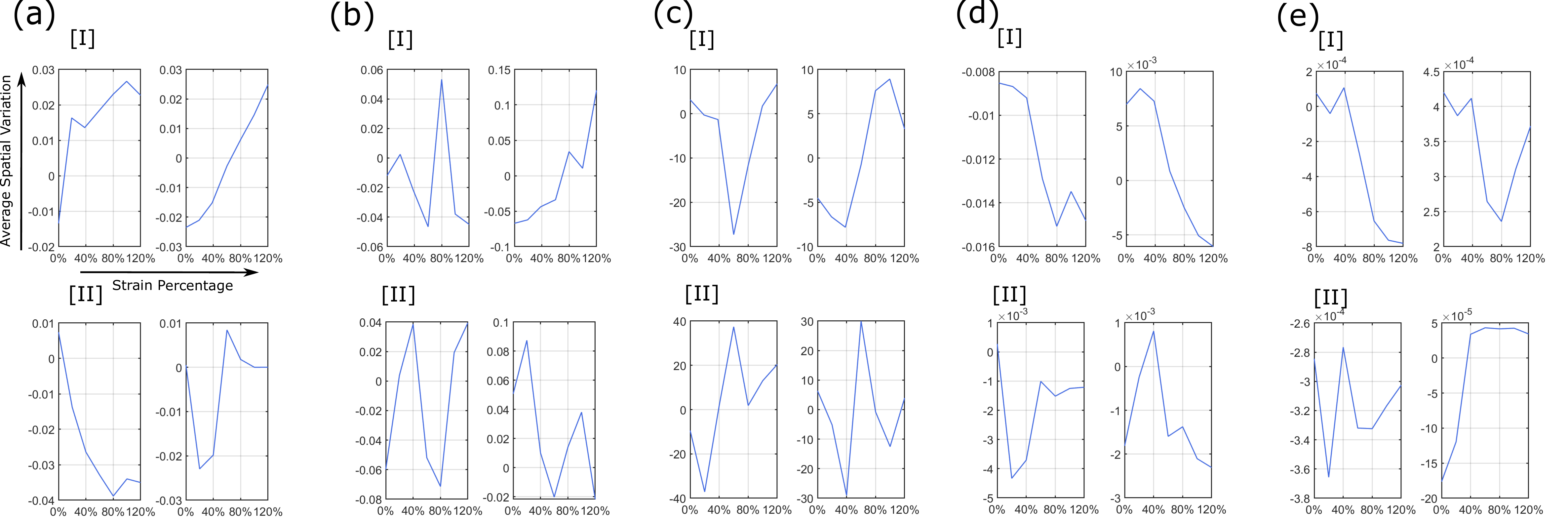}
     \caption{\textbf{Temporal changes in spatial variations of four representative quantities for rib fabrics, recorded across a tensile strain range of 0\% to 120\%.} (a)~Curve curvature. (b)~Curve torsion. (c)~Gaussian curvature. (d)~Area. (e)~Volume. Loading directions: [I]~{Course-wise}. [II]~{Wale-wise}. In each of (a)--(e), the left plot corresponds to {course-wise} variation and the right plot corresponds to {wale-wise} variation.}
     \label{fig5.SI.rib}
 \end{figure}

\begin{figure}[t!]
     \centering
     \includegraphics[width=1\linewidth]{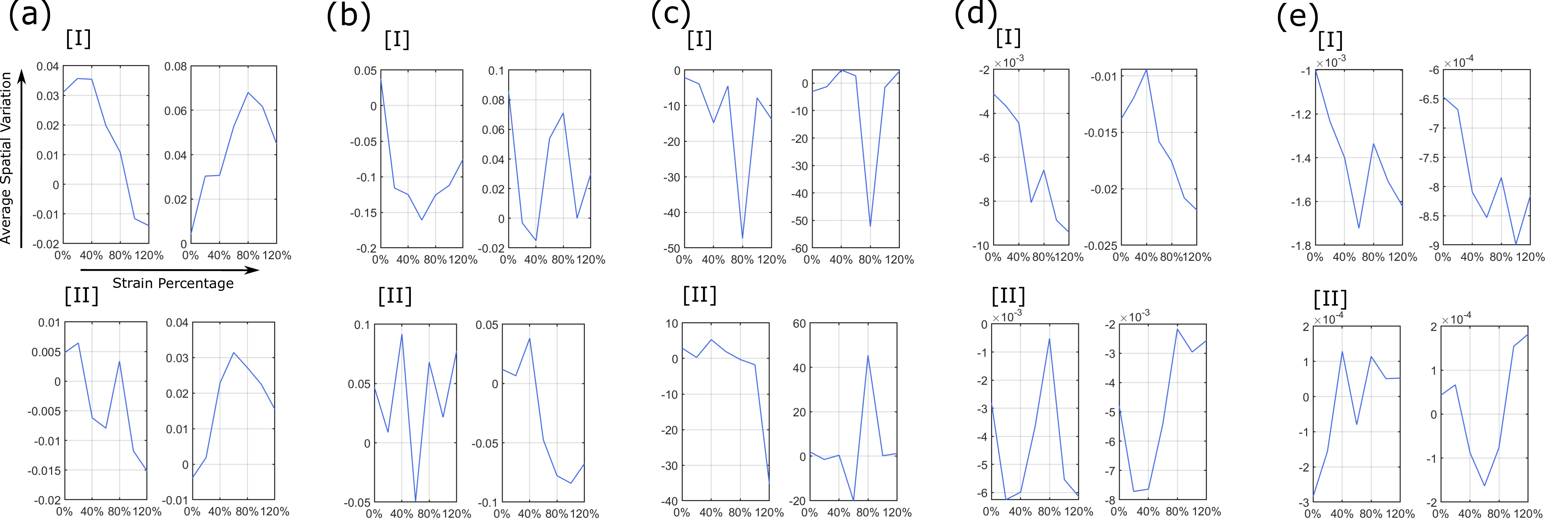}
     \caption{\textbf{Temporal changes in spatial variations of four representative quantities for seed fabrics, recorded across a tensile strain range of 0\% to 120\%.} (a)~Curve curvature. (b)~Curve torsion. (c)~Gaussian curvature. (d)~Area. (e)~Volume. Loading directions: [I]~{Course-wise}. [II]~{Wale-wise}. In each of (a)--(e), the left plot corresponds to {course-wise} variation and the right plot corresponds to {wale-wise} variation.}
     \label{fig5.SI.seed}
 \end{figure}

As for the temporal change of the spatial variation of 2.5D fabrics, besides the jersey results shown in the main text, here we further analyze the onset and evolution of hot spots in the garter, rib, and seed patterns (Fig.~\ref{fig5.SI.garter}--\ref{fig5.SI.seed}).

\end{document}